%% file: main.tex
\documentclass[sigplan]{acmart}
\renewcommand\footnotetextcopyrightpermission[1]{} 
\makeatletter                   
\pdfoutput=1
\def\mdseries@tt{m}             
\makeatother                    
\usepackage[plain]{fancyref}
\usepackage[draft=true]{minted} 
\usepackage{color}
\usepackage{hyperref}           
\hypersetup{
    colorlinks=true,
    linkcolor=blue,
    filecolor=red,      
    urlcolor=magenta,
    breaklinks=true,            
}
\usepackage{breakurl}           

\usepackage[normalem]{ulem}
\usepackage{amssymb}
\usepackage{amsmath}
\usepackage{caption}
\usepackage{subfig}
\usepackage[]{algorithm2e}
\usepackage{multirow}
\usepackage{graphicx}
\usepackage{mathtools}
\usepackage{physics}
\usepackage{physics}
\usepackage{listings}
\usepackage{minted}
\usepackage{multicol}
\usepackage{booktabs}
\usepackage{xspace}

\usepackage{tikz}
\usetikzlibrary{circuits.logic.US}
\usepackage{circuitikz}
\usepackage{qcircuit}

\usepackage{hyperref}
\newcommand{\tSMT} {{T-SMT}}
\newcommand{\tSMTstar} {{T-SMT$^\bigstar $}}

\newcommand{\rSMTstar} {{R-SMT$^\bigstar$\xspace}}
\newcommand{\greedyV} {{GreedyV$^\bigstar $}}
\newcommand{\greedyE} {{GreedyE$^\bigstar $}}

\newcommand{\Qiskit} {{Qiskit}}

\newcommand{\ibmnamefull} {{\em IBMQ 16 Rueschlikon}}
\newcommand{\ibmnameshort} {{\em IBMQ16}}

\newcommand{\hide}[1] {}

\setcopyright{none}

\begin{document}
\title{Noise-Adaptive Compiler Mappings \\ for Noisy Intermediate-Scale Quantum Computers}

\fancyhead[]{}

\author{Prakash Murali}
\affiliation{%
  \institution{Princeton University}
}
\authornote{Prakash Murali is the corresponding author and can be reached at pmurali@cs.princeton.edu.}
\author{Jonathan M. Baker}
\affiliation{%
  \institution{University of Chicago}
}

\author{Ali Javadi Abhari}
\affiliation{%
  \institution{IBM T. J. Watson Research Center}
}

\author{Frederic T. Chong}
\affiliation{%
  \institution{University of Chicago}
}

\author{Margaret Martonosi}
\affiliation{%
  \institution{Princeton University}
}

\renewcommand{\shortauthors}{P. Murali et al.}
\input{txt/abstract.tex}


\acmConference[Preprint of a publication in ASPLOS '19]{2019 Architectural Support for Programming Languages and Operating Systems}{April 13--17, 2019}{Providence, RI, USA}
\acmBooktitle{2019 Architectural Support for Programming Languages and Operating Systems (ASPLOS '19), April 13--17, 2019, Providence, RI, USA}
\acmPrice{15.00}
\acmDOI{10.1145/3297858.3304075}
\acmISBN{978-1-4503-6240-5/19/04}

\maketitle
\sloppy
\input{txt/introMRM.tex}

\input{txt/prelims.tex}
\input{txt/methodology.tex}
\input{txt/expt.tex}
\input{txt/results.tex}

\input{txt/related.tex}

\input{txt/conclusions.tex}

\begin{acks}
This work is funded in part by EPiQC, an NSF Expedition in
Computing, under grants CCF-1730449/1730082, in part by NSF PHY-1818914 and a research gift from Intel.
\end{acks}

\bibliographystyle{ACM-Reference-Format}

\end{document}

%% file: txt/abstract.tex
\begin{abstract}
A massive gap exists between current quantum computing (QC) prototypes, and the size and scale required for many proposed QC algorithms.  Current QC implementations are prone to noise and variability which affect their reliability, and yet with less than 80 quantum bits (qubits) total, they are too resource-constrained to implement error correction.  The term Noisy Intermediate-Scale Quantum (NISQ) refers to these current and near-term systems of 1000 qubits or less.  Given NISQ's severe resource constraints, low reliability, and high variability in physical characteristics such as coherence time or error rates, it is of pressing importance to map computations onto them in ways that use resources efficiently and maximize the likelihood of successful runs.  

This paper proposes and evaluates backend compiler approaches to map and optimize high-level QC programs to execute with high reliability on NISQ systems with diverse hardware characteristics. Our techniques all start from  an LLVM intermediate representation of the quantum program (such as would be generated from high-level QC languages like Scaffold) and generate QC executables runnable on the IBM Q public QC machine. We then use this framework to implement and evaluate several optimal and heuristic mapping methods.  These methods vary in how they account for the availability of dynamic machine calibration data, the relative importance of various noise parameters, the different possible routing strategies, and the relative importance of compile-time scalability versus runtime success. 
Using real-system measurements, we show that fine grained spatial and temporal variations in hardware parameters can be exploited to obtain an average $2.9$x (and up to $18$x) improvement in program success rate over the industry standard IBM Qiskit compiler. Despite small qubit counts, NISQ systems will soon be large enough to demonstrate ``quantum supremacy,'' i.e., an advantage over classical computing.  Tools like ours provide significant improvements in program reliability and execution time, and offer high leverage in accelerating progress towards quantum supremacy. 
\end{abstract}

%% file: txt/introMRM.tex
\section{Introduction}
Quantum computing (QC) aims to solve intractable computational problems by leveraging quantum mechanical principles like superposition and entanglement to manipulate information efficiently. QC algorithms show potential to significantly impact areas such as quantum chemistry \cite{vqe1, vqe2}, cryptography \cite{shor1}, machine learning \cite{quantum_ml1}, and others.  Unfortunately, a massive gap exists between the resources required by most proposed QC algorithms, and the resources which exist in current prototype hardware.  

 QC systems have been announced with 49-72 qubits \cite{googlebristlecone,ibm50q,intelq} and current operational systems have been demonstrated publicly with roughly 20 qubits or fewer \cite{ibmq}.  A QC system with 72 fully-entangled qubits and sufficiently-precise operations (``gates'') would likely be sufficient to show ``quantum advantage'' over the largest classical supercomputers, but would still be 5-6 orders of magnitude smaller than the resource requirements of Shor's well-known QC algorithm for factoring large numbers \cite{shor1, Devitt2013, shor_est1}.  

The term Noisy Intermediate-Scale Quantum (NISQ) computers refers to  the current and near-term QC systems which have roughly 1000 qubits or fewer---typically too small to employ error correction codes (ECC) \cite{nisq}.  While resource constrained, NISQ machines offer an important step forward: if used well, they can demonstrate QC applications generating useful results.  Making good use of NISQ hardware, however, requires very efficient, near-optimal mappings of algorithms onto them. This paper proposes a suite of optimization- and heuristic-based approaches for mapping applications onto NISQ hardware, and evaluates them by running the mapped executables on a public 16-qubit IBM system\footnote{We run all experiments on the 16-qubit IBM instance named \ibmnamefull\ \cite{ibmq}. For the remainder of the paper, we shorten this name to \ibmnameshort.}.  

A good mapping of a QC algorithm onto NISQ hardware requires first an intelligent initial placement of the program qubits onto the hardware qubits in order to reduce communication requirements. Second, it requires efficient orchestration of operations both for the computation itself, and also for the additional SWAP operations which communicate state between hardware qubits. Third and most importantly, mapping decisions must reduce the likelihood of operational or decoherence errors which cause the program run to fail to achieve a useful answer. Our work performs mappings using the daily calibration data provided by IBM in order to avoid using unreliable qubits and to prioritize qubit positioning which reduces the likelihood of communication (SWAP) errors.  For example, Figure \ref{fig:hw_var2} shows large daily variations in the gate error rates and coherence times of the qubits of \ibmnameshort\ on which we experiment. Our contributions are:

\begin{figure}[t]
\centering
\subfloat[Coherence time (T2)]
{
    \includegraphics[scale=0.5]{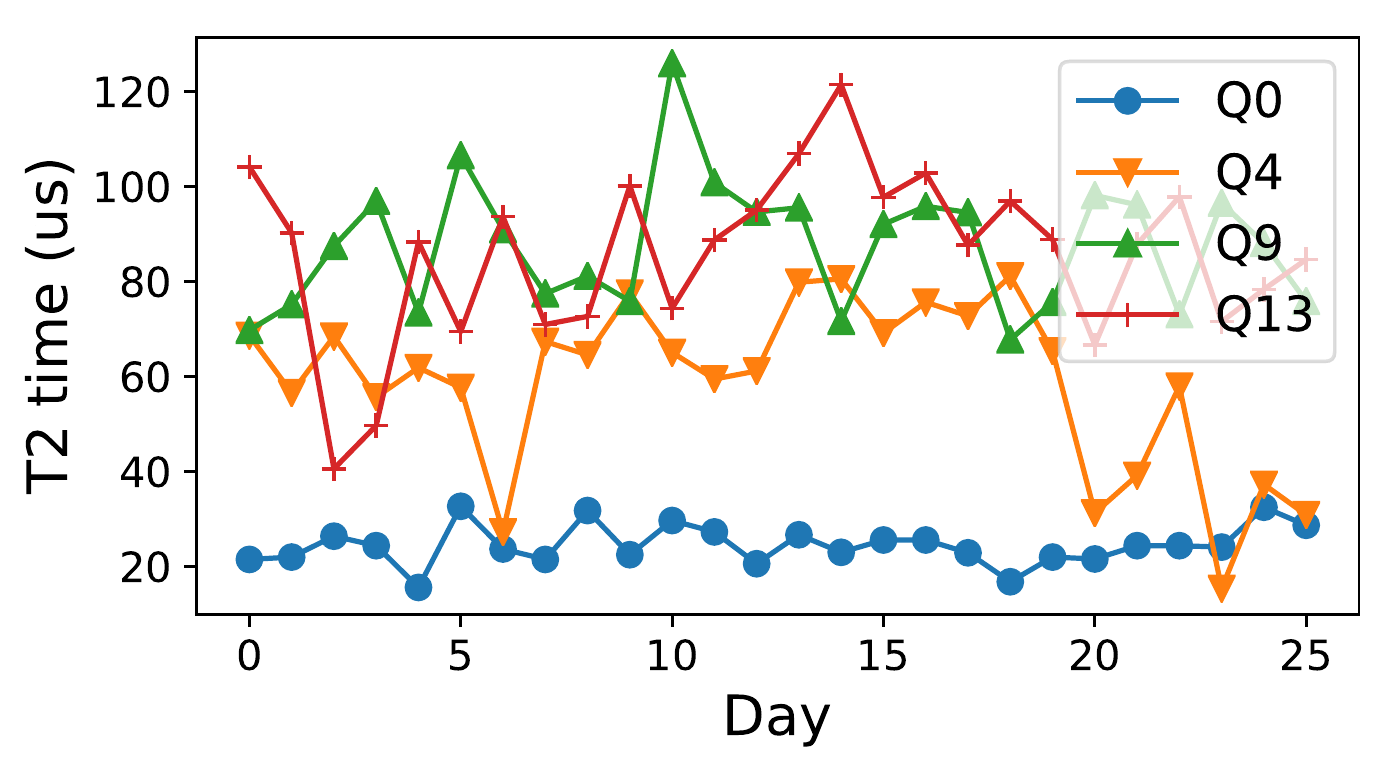}
    \label{fig:ibmqx5_coherence}
}
\vspace{-1em}
\subfloat[CNOT gate error rate]
{
    \includegraphics[scale=.5]{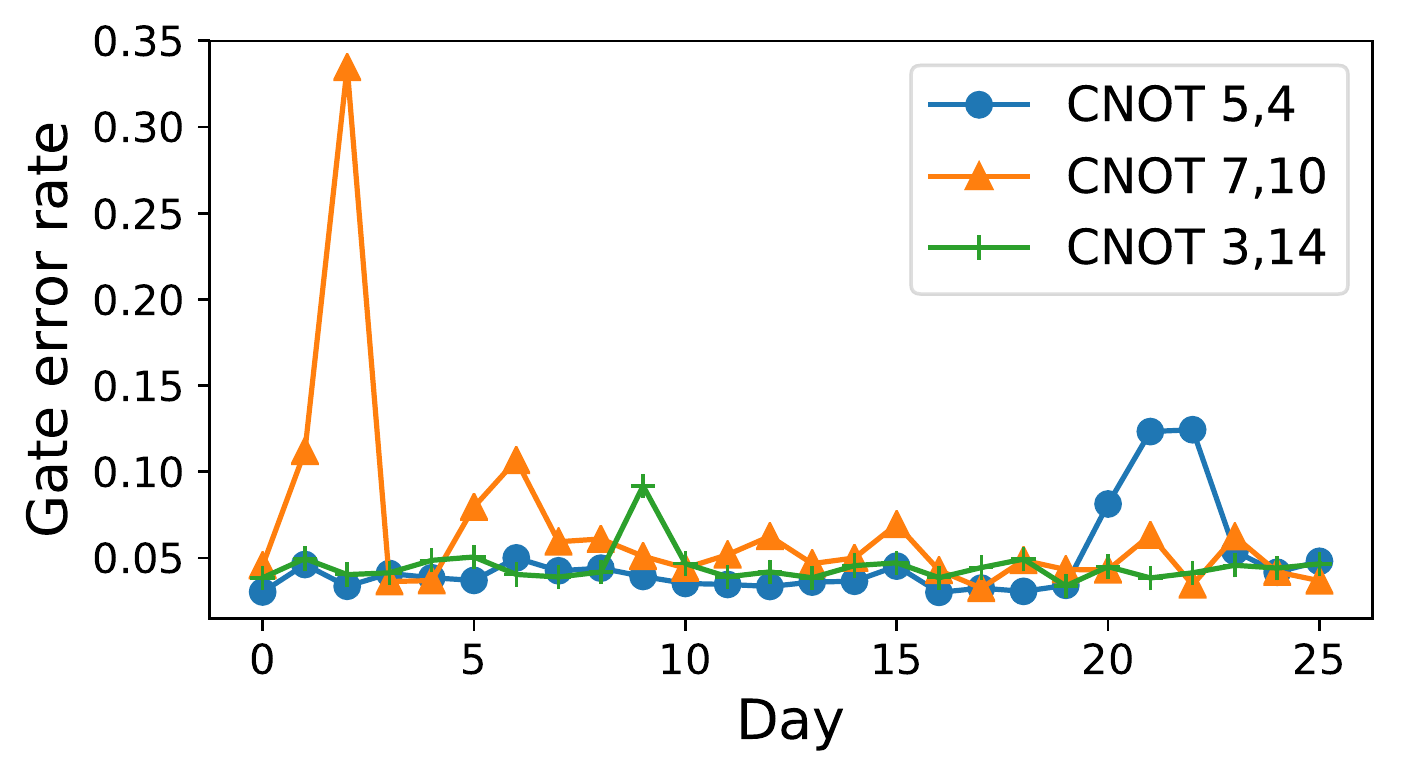}
    \label{fig:ibmqx5_cx_error}
}
\caption{Daily variations in qubit coherence time (larger is better) and gate error rates (lower is better) for selected elements in \ibmnamefull. The qubits and gates that are most or least reliable are different across days.}
\label{fig:hw_var2}
\end{figure}

First, we develop an LLVM \cite{llvm} compiler which optimally or near-optimally maps quantum programs to OpenQASM assembly code \cite{openqasm1} and then to the web-accessible \ibmnameshort\ machine for real-system evaluation. For 12 QC programs written in the Scaffold quantum programming language \cite{scaffold}, we use this framework to explore how optimal and heuristic mapping methods, qubit movement policies, and the intelligent adaptation to machine calibration data can affect the quality of the compiled code.

In particular, our compiler provides up to $1.68$x gain in execution time and $9$x gain in success rate over an optimal but calibration-unaware baseline. Our compiler obtains an average $2.9$x improvement (up to $18$x) in success rate, and an average $2.7$x improvement in execution time (up to $6$x), compared to the IBM Qiskit compiler \cite{qiskit}, which is the industry standard for \ibmnameshort.

Furthermore, although compile-time is not a first-order design goal, QC compilers must scale well enough for intelligent compilation to be tractable throughout NISQ-range machines.  We show that our methods based on Satisfiability Modulo Theory (SMT) scale well up to 32 qubits.  Further, we have developed calibration-aware heuristic methods which produce executables with similar reliability and execution time as the SMT approaches, but with more scalable compile-times beyond 32 qubits.

Finally, across the 12 benchmarks, we study the influence of application instruction mix and time varying qubit error characteristics on compiled programs. For example, applications for which our compiler can identify zero-qubit-movement mappings have substantially higher likelihood of success (up to $2.8$x), compared to programs which require even a single qubit movement operation.

Overall, NISQ systems are important to QC progress because their success in demonstrating quantum supremacy and running small but useful QC programs is an important stepping-stone in the maturation of this technology.  In its leveraging of intelligent and calibration-aware mapping techniques to significantly improve execution time and success rate of quantum executions, our tool makes an important contribution in helping close the gap to quantum supremacy and advancing toward practical QC.


%% file: txt/prelims.tex

\section{Background on Quantum Computing}\label{sec:background}

\hide{
\begin{figure}[t]
    \centering
    \includegraphics[scale=0.45]{figs/bv43.pdf}
    \caption{Example of quantum circuit for the Bernstein-Vazirani on 4 qubits with hidden string $s = 111$.}
    \label{fig:bv4_3}
\end{figure}}

{\noindent \bf Principles of Quantum Computing: } A qubit is the basic unit of quantum information. 
Unlike classical bits, which take two values (0 and 1), superposition allows qubits to be in a probabilistic combination of the two states.
If we consider the states $\ket{0}$ and $\ket{1}$ as basis vectors of $\mathbb{C}^2$, we can express the state of a qubit $\ket{\psi}$ as $\ket{\psi} = \alpha\ket{0} + \beta\ket{1}$, where $\alpha$ and $\beta$ are complex amplitudes such that $|\alpha|^2 + |\beta|^2 = 1$. 
The state of one or more qubits can be manipulated by modifying the complex amplitudes using operations termed as gates. Single-qubit operations include: H, X, Y, Z and others. The act of measurement or readout collapses the superposition state to one of the two basis vectors, a classical output.

A controlled NOT (CNOT) gate is an example of a two-qubit gate. A CNOT gate has a control and target qubit. When the control qubit is in the state $\ket{1}$, the state of the target bit is flipped. In quantum CNOT gates, the gate can operate on qubits to 
entangle them to have non-classical correlations in their states and measurement outputs. We use the notation {\tt CNOT C, T} for a CNOT gate with control C and target T. 

A quantum computer with $n$ fully-entangled qubits has an exponential state space of size $2^n$. In a QC application, a set of qubits are initialized to encode a given problem including its data input.  As the program executes, qubit amplitudes are manipulated, typically to boost the probabilities of the desired outcomes in the state space. Finally, the qubits are measured to produce classical output for the given problem.

{\noindent \bf NISQ Systems: } NISQ systems are near-term quantum systems expected to scale to a few hundred qubits, paving the way towards large-scale QC \cite{nisq}. 
Qubits in NISQ systems have short coherence time, high gate error rates and and limited qubit connectivity. They are typically too resource-constrained to implement error-correcting codes (ECC).

\begin{figure*}[t]
\centering
\subfloat[Bernstein-Vazirani Intermediate Representation]
{
    \includegraphics[scale=0.35]{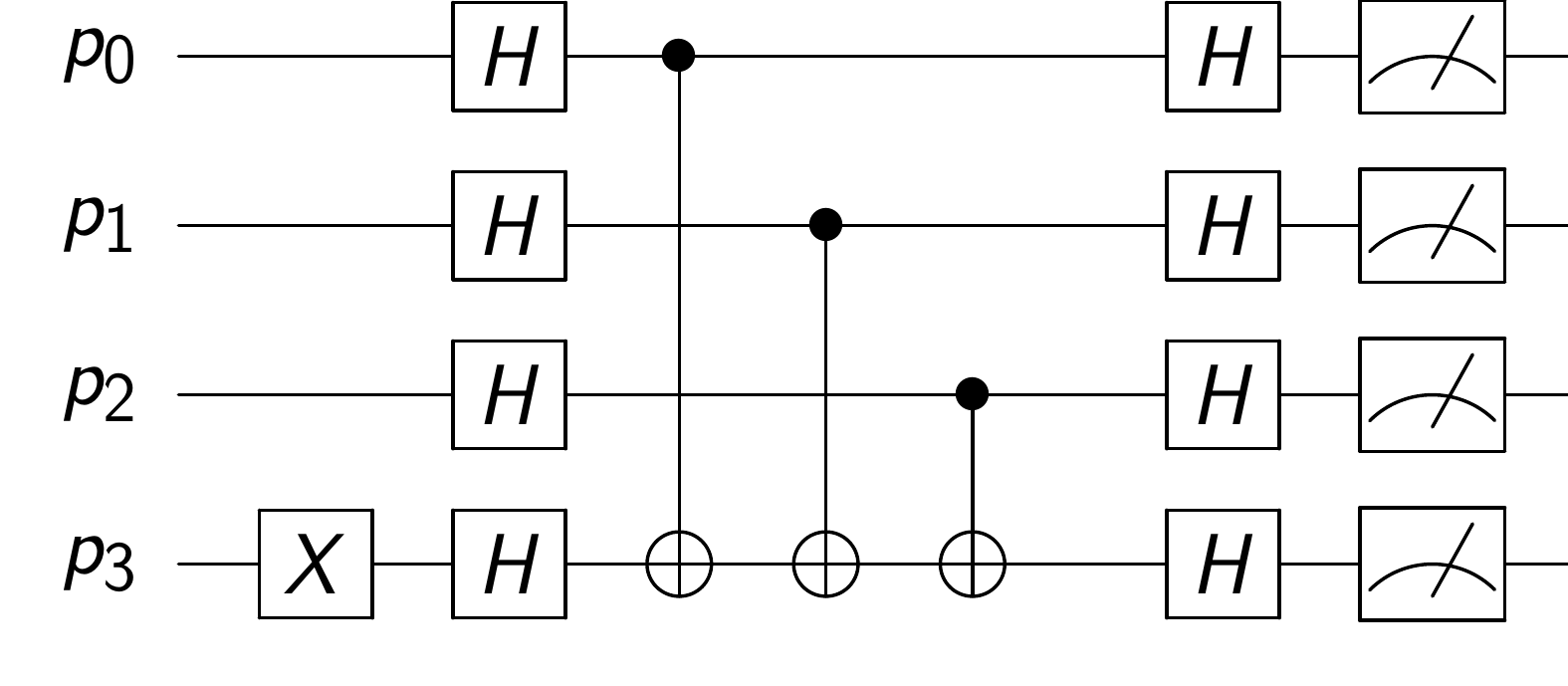}
    \label{fig:bv4_code}
}
\qquad
\subfloat[Layout of qubits in \ibmnameshort\ and a naive mapping for BV4.]
{
    \includegraphics[scale=.45]{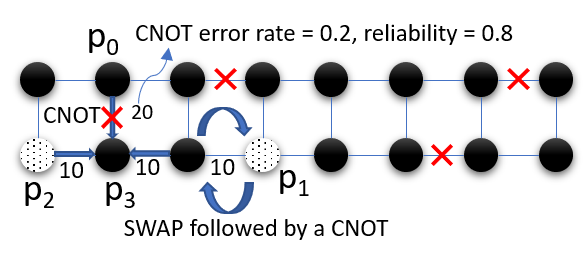}
    \label{fig:bv4_random}
}
\qquad
\subfloat[Optimized mapping for BV4]
{
    \includegraphics[scale=.4]{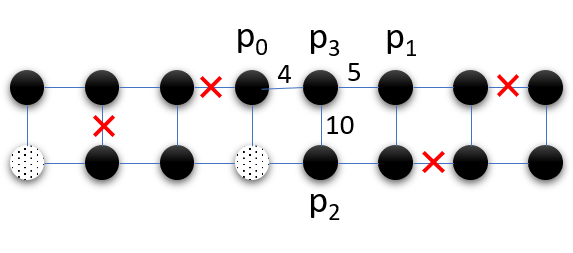}
    \label{fig:bv4_goodmap}
}
\label{fig:bv_mappings}
\caption{Figure (a) shows the intermediate representation of the Bernstein-Vazirani algorithm on 4 qubits (BV4). Each program qubit is represented by a line. X and H are single qubit gates. The CNOT gates from each qubit $p_{0,1,2}$ to $p_3$ are marked by vertical connectors. The measurement or readout operation is indicated by the meter. Figure (b) shows the layout of the hardware qubits in \ibmnameshort\ and a naive mapping of BV4's program qubits. The black circles denote qubits and the edges indicate permitted CNOT gates. The numbers on the labelled edges indicate the CNOT gate error ($\cross10^{-2}$). The hatched qubits and crossed gates are unreliable. In this mapping, qubit movement is required to perform the CNOTs and error-prone operations are used. Figure (c) shows a mapping where qubit movement is not required and unreliable qubits and gates are avoided.}
\end{figure*}

As a concrete NISQ example, Figure \ref{fig:bv4_random} shows the layout of the qubits in the 16-qubit IBM system. This system implements a set of 1- and 2-qubit operations, akin to an instruction set.  For 2-qubit operations, this machine only supports hardware CNOT gates being performed  between {\em adjacent} qubits, based on the topology shown in Figure \ref{fig:bv4_random}. To perform CNOT gates between non-adjacent qubits, we should use SWAP operations between adjacent qubits until the two of interest for a given CNOT computation are in adjacent locations. Each SWAP operation between two adjacent qubits itself requires 3 CNOT gates\footnote{For two qubits $X$ and $Y$, {\tt SWAP(X,Y) :=  \{CNOT X,Y; CNOT Y,X; CNOT X,Y\}} \cite{Mermin}.}  Our compiler aims to reduce the {\em time cost} of these operations.  More importantly, each one of these operations incurs some error, so a key goal of our optimization is to reduce operation counts and error rates in order to increase the likelihood of an overall successful run. We refer to this as {\em reliability} and it is the primary design goal of this work.

In addition to compiler optimization based on attributes like gate counts, our approach also adapts based on publicly-available experimental data.  In particular, the IBM Q machines are calibrated twice a day. Once a day there are public postings of experimental measurements of key properties: qubit relaxation time (T1), coherence time (T2), gate errors and readout errors \cite{ibmqexp}. From daily calibration logs, we observe that qubit coherence time is $70$ microseconds on average, but varies up to $9.2$x spatially and temporally across qubits and daily calibrations. The average error rate for CNOTs is $0.04$, readouts is $0.07$ and single qubit gates is $0.002$. CNOT and readout error rates exhibit up to $9.0$x and $5.9$x variation across qubits and calibration cycles, respectively. CNOT gate durations vary up to $1.8$x across qubits. These fluctuations stem from material defects caused by the lithographic processes used to manufacture the qubits and are expected to be present in future generations of superconducting qubits also \cite{superconducting_stability}. 
 
These error rates imply only very short programs can execute reliably on the machine. A program with more than 16 CNOT operations, has less than $50\%$ chance of executing correctly.  A key goal of our compiler optimizations is to use this calibration data to boost the success rate of individual program runs, by avoiding portions of the machine with poor coherence, operation, or readout errors.

%% file: txt/methodology.tex
\section{Compilation Framework: Overview}\label{sec:overview}

Our framework takes a Scaffold program \cite{scaffold} as input, and produces compiled OpenQASM code \cite{openqasm1}. The Scaffold quantum programming language extends C with quantum gates. Scaffold programs are independent of the machine topology, size and qubit properties. The ScaffCC compiler \cite{scaffcc1, scaffcc2} performs automatic gate and rotation decomposition, implements high level operations like the Toffoli gate and produces an LLVM Intermediate Representation (IR) \cite{llvm} of the program. The IR version of the program includes the qubits required for each operation and the data dependencies between operations. For example, Figure \ref{fig:bv4_code} shows the IR for the simple 4-qubit Bernstein-Vazirani algorithm which is chosen because it fits on machines of this size and has an answer which can be calculated to check our results \cite{bernsteinvazirani}. We use the program IR as a starting point for the noise-aware backend described here. 

Starting from the IR, the noise-aware backend has three primary tasks. First, qubits in the program must be {\bf mapped} to distinct qubits in the hardware implementation, preferably in a way that reduces qubit state movement required as the program executes. Second, the compiler performs {\bf operation scheduling} while respecting data dependencies between gates. To accomplish this, each operation is assigned a start time constraint, and the scheduler emits control code that enforces this. Third, to perform 2-qubit operations on non-adjacent qubits, the compiler should orchestrate {\bf communication through SWAPs}. That is, it automatically inserts the required SWAP operations to bring the qubits adjacent to each other before the operation is performed. 

\begin{figure}
    \centering
\includegraphics[scale=0.52]{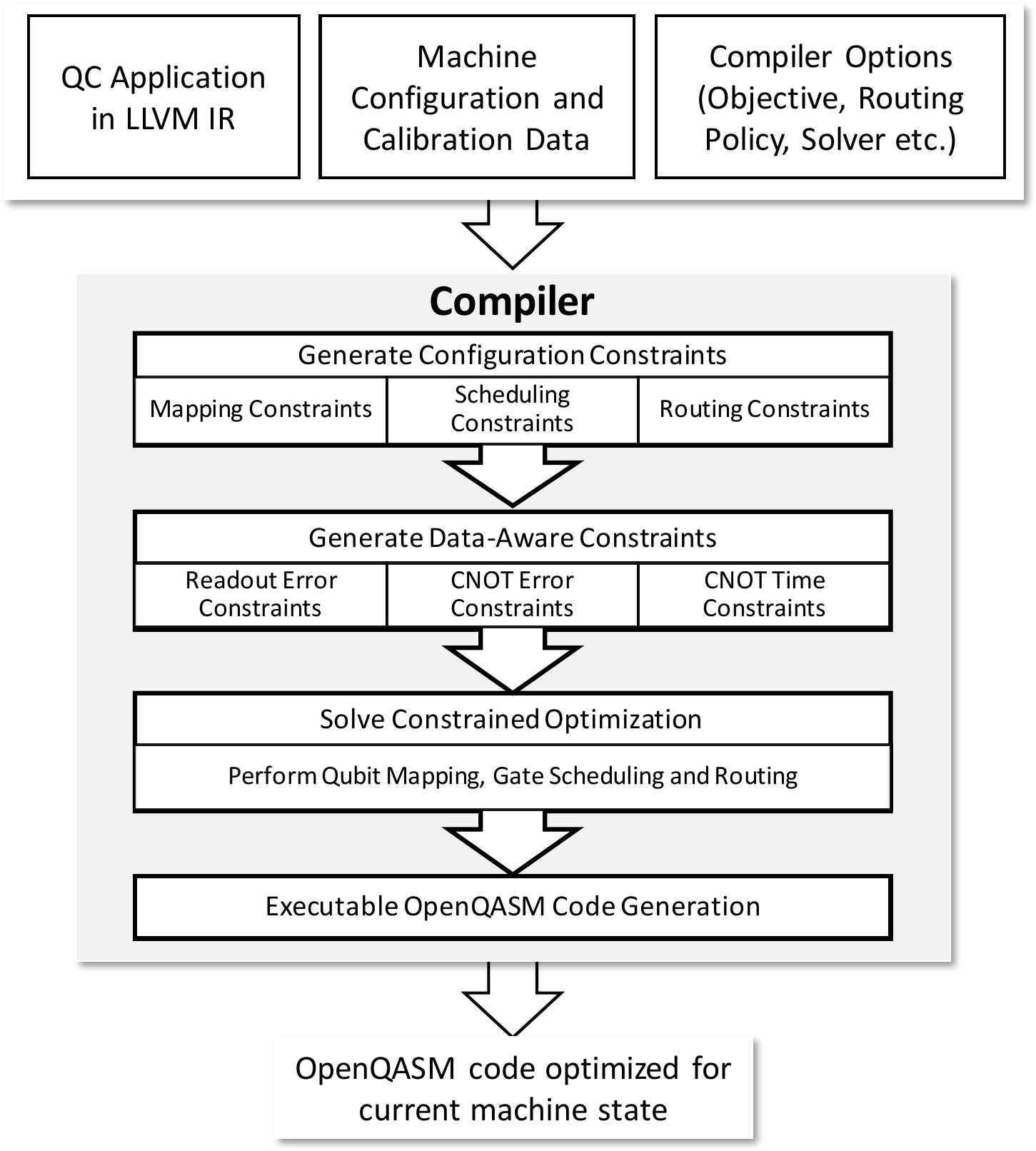}
    \caption{Optimization Pipeline. Inputs are a QC program, details about the specific hardware configuration, and a set of options, such as routing policy and solver approach. From these, compiler generates a set of appropriate constraints and uses them to map program qubits to hardware qubits and schedule operations. Finally, the compiler generates an executable version of the program, here for \ibmnameshort.}
    \label{fig:pipeline}
\end{figure}

Consider a simple compilation method where program qubits are assigned to random qubits on the hardware. Figure \ref{fig:bv4_random} shows such a mapping for the BV4 IR. In this mapping, the compiler must insert qubit movement or swap operations to perform the CNOT gates between $p_{1}$ and $p_3$. In contrast, the mapping shown in Figure \ref{fig:bv4_goodmap} requires no qubit movement because the qubits required for the CNOTs are adjacent. In addition, this mapping is noise-aware; namely, it uses the calibration data to select a mapping that avoids using qubits with low coherence time and gates with high error rates. Our compiler uses machine topology and calibration data to automatically generate such mappings for a given program.

Our primary goal is to maximize the likelihood that the program runs successfully. To accomplish this, we have three main strategies. First, the compiler places program qubits on hardware locations with high reliability, based on the calibration data. The compiler considers the effect of errors due to CNOTs and readouts; for this machine, single-qubit error rates are considerably smaller so our formulation chooses to ignore them. Second, to mitigate errors due to decoherence, the compiler should schedule all gates to finish before the coherence time of the hardware qubits (intuitively analogous to making use of data within the refresh interval of a DRAM). Third, the compiler optimizes for the qubit topology to avoid unnecessary qubit movement. Qubit movement not only increases execution duration, but more importantly leads to high error rates since each qubit SWAP operation includes three error-prone CNOTs. We have designed a set of optimal and heuristic compilation variants to accomplish these goals. 

\begin{table*}[t]
\centering
\small
\begin{tabular}{|c|c|c|c|}
\hline
Algorithm & Objective & Parameters & Constraints \\                                   
\hline \hline
\Qiskit   &  Heuristic, minimize duration  & - & -  \\ \hline
\tSMT     & SMT solver, minimize duration & Routing policy: RR, 1BP & 1-4, 7-9 \\ \hline
\tSMTstar &  SMT solver, minimize duration & Routing policy: RR, 1BP & 1-3, 5-9 \\ \hline
\rSMTstar    & SMT solver, maximize reliability  & \begin{tabular}[c]{@{}l@{}}Routing policy: 1BP\\ Readout weight $\omega \in [0,1]$\end{tabular} & 1-3, 5-6, 9, 10-11\\ \hline
\greedyV  & Heuristic, maximize reliability&  Routing Policy: Best Path & - \\ \hline
\greedyE   & Heuristic, maximize reliability & Routing Policy: Best Path & -  \\ \hline

\end{tabular}
\caption{List of compiler configurations used in our study. The IBM Qiskit 0.5.7 compiler is used as a the baseline. The use of calibration data is marked by a $\bigstar$.}    
\label{tab:compiler_config}
\end{table*}
    
Table \ref{tab:compiler_config} enumerates the full set of compiler variants we consider in this paper.  In addition to the publicly-available IBM Qiskit compiler we use as a comparative baseline, we also develop several approaches which are either truly optimization-based or heuristic. We give an overview of these approaches here, before offering details in the following section.
    
\subsection{Optimization-Based Mappings}
 
In the optimization-based variants of our compiler, we implement the above goals by posing the compilation problem as a constrained optimization problem to be solved by a satisfiability modulo theory (SMT) solver. The optimization problem has variables and constraints which express program information, machine topology constraints, and machine error information. The variables include program qubit locations, gate start times and routing paths. The constraints specify qubit mappings should be distinct, gates should start in program dependency order, and routing paths should be non-overlapping. Fig. \ref{fig:pipeline} summarizes the general compilation pipeline for the solver-based approach, beginning with an IR of a program and resulting in execution-ready code.

The optimization objective is to maximize the reliability or success rate of program runs. We express the reliability of the program as the product of the reliability of all gates in the program. (Because of the degree of entanglement in QC programs, this serves as a useful measure of overall correctness.) For a given mapping, the solver determines the reliability of each program CNOT, readout operation and single qubit gate and computes an overall reliability score. For the optimization variants which are noise-aware, the solver can maximize the reliability score over all mappings by tracking and adapting to the error rates, coherence limits, and qubit movement based on program qubit locations.

\hide{
\begin{figure}[t]
    \centering
    \includegraphics[scale=0.6]{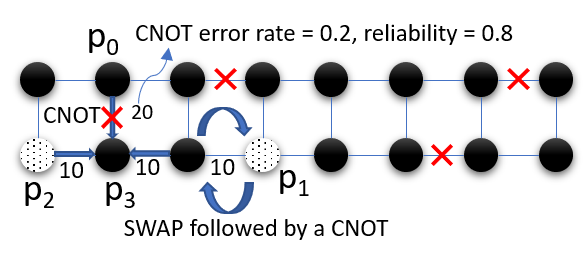}
    \caption{An example mapping for the BV4 IR. The numbers on the edges indicate CNOT gate error rates. To perform the CNOT gate between $p_1$ and $p_3$, $p_1$ has to SWAP once with its left neighbor and become adjacent to $p_3$.}
    \label{fig:bv4_reliability}
\end{figure}
}

Given a target machine, our framework converts the program IR into an optimization problem by expressing an objective and constraints that can be solved using an Satisfiability Modulo Theory (SMT) solver \cite{z3, omt_z3}. For classical programs, these solvers have been used to obtain optimal hardware mapping and scheduling for spatial architectures \cite{Nowatzki:2013:GCS:2491956.2462163}, but to our knowledge, ours is the first use of them for QC systems. SMT solvers take as input a set of linear constraints, and an objective function and search for an optimal solution. Although the reliability objective is a product of individual gate reliability scores (and therefore non-linear), we linearize the objective by instead optimizing for the additive logarithms of the reliability scores. An SMT solver can then be invoked to find a mapping which maximizes the log reliability. 

{\noindent \bf Does maximizing the reliability score achieve our goal of increasing program success rate?} Optimizing for the reliability score induces the compiler to place qubits at locations where CNOT and readout errors are low. It also indirectly minimizes qubit movement because CNOTs between far away qubits are error-prone. For example, for the BV4 IR, consider mapping shown in Figure \ref{fig:bv4_random}. Here, the reliability of the CNOT between $p_0$ and $p_3$ is $0.8$ ($80\%$ chance of executing correctly), while the reliability of the CNOT between $p_1$ and $p_3$ is only $0.65$\footnote{$p_1$ has to swap once to move to a location adjacent to $p_3$. The net reliability of the 3 CNOTs required to perform the SWAP is $0.9^3 = 0.729$. Then the actual CNOT operation can be performed with reliability 0.9. Hence, the overall CNOT reliability is $0.65$.}. Thus, the compiler will choose mappings where communicating qubits are close together, minimizing unnecessary qubit movement and allowing gates to be scheduled to finish within the coherence window.

\subsection{Heuristic Mappings}
We also determine whether heuristic techniques can approach the optimization-based results, but with better scalability.  For this, we develop two comparative algorithms based on greedy heuristics.  The greedy heuristics analyze the CNOTs in the program IR, and determine a gate frequency for each qubit and program CNOT. 

We explore two policies. In the first policy, \greedyV, we place program qubits on hardware qubits in the heaviest qubit first order.  In the second policy, \greedyE, we place program CNOTs and their control and target qubits in a heaviest edge first order. 
Intuitively, the first policy places qubits which use more CNOTs in locations which have good CNOT and readout error rates. The second policy places pairs of qubits which have the most frequent CNOTs first. 

\section{Optimal Compilation}\label{sec:constraints}

\subsection{Notations and Assumptions}
Let $Q_P$ be the set of program qubits. Let $Q_H$ be the set of hardware qubits. In this work, we assume hardware qubits are arranged as a 2-D grid of dimensions $M_x \times M_y$. Likewise, due to the connectivity characteristics of \ibmnameshort, we assume only hardware qubits which are adjacent in the grid are permitted to participate in two qubit operations. More  elaborate topology and routing assumptions can be handled in future work.  For $q \in Q_P$, the ordered pair $(q.x, q.y)$ corresponds to the location of the hardware qubit assigned to the program qubit $q$. 
Let $G$ be the set of operations in the program. This includes single-qubit gates such as $H$, and the 2-qubit $CNOT$ gate and qubit measurement or $Readout$ operations. CNOT and readout operations dominate the reliability outcomes, so the reliability score focuses on them.  The subset of CNOT gates is denoted by $G_{CNOT}$, and the subset of readout  operations is $G_{Readout}$. 
For each gate $g$ in the program, the start time is denoted by ($g.\tau$), duration by ($g.\delta$), and reliability by ($g.\epsilon$).
To denote data dependencies between the operations, we use a binary relation $>$ on the gates, so that for two operations $g_2 > g_1$ if $g_2$ depends on $g_1$. Although the reliability objective focuses on a subset of operations, we map and schedule all operations (including single-qubit operations) to provide a valid real-system executable.  


\subsection{Constraints}

{\noindent \bf Qubit Mapping Constraints:} Constraint \ref{c1}, guarantees all program qubits are mapped to actual hardware qubits. Constraint \ref{c2} guarantees each program qubit is assigned a unique location.
\begin{align}
    \forall q \in Q_P : 0 \le q.x < M_x \land 0 \le q.y < M_y \label{c1}\\
    \forall q_1, q_2 \in Q_P : q_1.x \neq q_2.x \lor q_1.y \neq q_2.y \label{c2}
\end{align}

{\noindent \bf Gate Scheduling Constraints: }
For each gate $g$ in the program, the compiler determines the start time and execution duration. If two gates $g_1$ and $g_2$ both operate on the same qubit, and $g_2$ uses the output of $g_1$, $g_2$ should start only after $g_1$ finishes. For every such edge in the dependency graph, Constraint \ref{c4} shows the form we use to enforce such data dependencies.
\begin{gather}
    \forall g_1, g_2 \in G : g_2 > g_1 \Rightarrow g_2.\tau \ge g_1.\tau + g_1.\delta \label{c4}
\end{gather}

The durations, $\delta$, for single qubit operations are set using the documented durations in timeslots of the corresponding hardware operations. For CNOTs, the duration includes both the operation itself as well as the time to bring the relevant program qubit states into adjacent hardware qubits; this depends on the routing policy and is discussed below. 

{\noindent \bf CNOT Duration based on Grid Distance: } The duration of a CNOT gate accounts for both CNOT time and the duration of the swap paths before and after the CNOT. For a CNOT $g \in G_{CNOT}$, let the control and target qubits be $q_c$ and $q_t$. Then the duration of the CNOT is: $g.\delta = 2*(\norm{q_c - q_t}_1-1)*\tau_{SWAP} + \tau_{CNOT}$ where $\norm{q_c - q_t}_1 = \abs{q_c.x - q_t.x} + \abs{q_c.y - q_t.y} \label{c5}$ and $\tau_{SWAP}$, $\tau_{CNOT}$ are the times to complete a SWAP or CNOT operation, respectively. 

The compiler must schedule operations before the individual qubits decohere. For \tSMT\ (noise-unaware) we simply use an assumption of $M_T$ as 1000 timeslots of coherence time, which is the long-term average for the machine:

\begin{align}
\forall g \in G : g.\tau + g.\delta < M_T \label{c3}
\end{align}

{\noindent \bf CNOT Duration based on Calibration Data: } For \tSMTstar\ and \rSMTstar, we set durations based on calibration data.  In particular, since qubit coherence time changes daily (Figure \ref{fig:ibmqx5_coherence}) and CNOT gate durations vary across qubits, these approaches use the calibration-based data in the optimization constraint. 
To set durations based on calibration data, we assume a routing policy and compute the CNOT durations for each hardware qubit pair. Let $\Delta$ be an $|Q_H|\times|Q_H|$ matrix where $\Delta_{h_i, h_j}$, $i\neq j$, specifies the duration of a CNOT between hardware qubits $h_i, h_j \in Q_H$. The duration of a program CNOT can be set as: for all $g \in G_{CNOT}$ and for all $h_1, h_2 \in Q_H$:
\begin{align}
    g_c = h_1 \land g_t = h_2 \Rightarrow g.\delta = \Delta_{h_1, h_2} \label{dyn-duration-const}
\end{align}

For the calibration-aware coherence time bound, constraint \ref{dyn-coherence-const} ensures every gate finishes before the coherence time of the qubits it acts on i.e., if a gate uses a hardware qubit $h$, it should complete before $h$ decoheres, with $h.\tau$ as the coherence time of a hardware qubit $h \in Q_H$. We have for all $g \in G$ and for all $h_1, h_2 \in Q_H$: 
\begin{align}
    g_c = h_1 \land g_t = g_2 \Rightarrow g.\tau + g.\delta \le \min{(h_1.\tau, h_2.\tau)} \label{dyn-coherence-const}
\end{align}

\subsection{Routing for CNOT Gates}\label{sec:43}

\begin{figure}
\centering

\subfloat[Rectangle Reservation (RR)]
{
    \includegraphics[scale=.4]{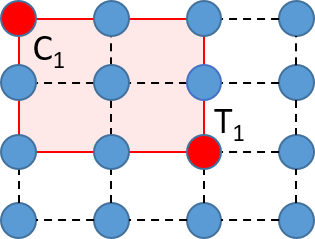}
    \label{figRR}
}
\quad
\subfloat[One Bend Paths (1BP)]
{
    \includegraphics[scale=.4]{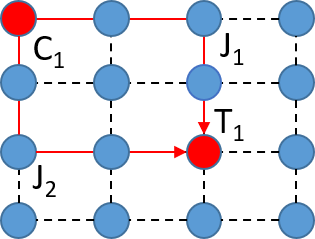}
    \label{figOBP}
}
\caption{Two routing policies for swap-based architectures.}
\label{fig:routing}
\end{figure}

To route multiple CNOTs in parallel, the compiler uses two routing policies based on policies in VLSI routing \cite{gonzalez, routing1}.

{\noindent \bf Rectangle Reservation:}
In this policy, for every CNOT in the program, the compiler blocks a 2D region bounded by the control and target qubit, during the CNOT execution. For example, in Figure \ref{figRR}, the highlighted rectangle is reserved for the duration of the CNOT.

Consider a CNOT gate $g_i \in G_{CNOT}$. Let $(l_x^i, l_y^i)$ and $(r_x^i, r_y^i)$ denote the top left and bottom right corners, respectively, of the bounding rectangle of $g_i$. These variables are defined using min and max relations on the qubit mapping variables of the CNOT. For two CNOTs $g_i$ and $g_j$, the routing constraint is:
\begin{align}
S(R_i,R_j) = \lnot(l_x^i > r_x^j \lor  r_x^i < l_x^j \lor l_y^i > r_y^j \lor  r_y^i < l_y^j) \\  
T(g_i,g_j) = \lnot(g_i.\tau > g_j.\tau + g_j.\delta \lor g_j.\tau > g_i.\tau + g_i.\delta)
\end{align}
Constraint $S$ checks if the two rectangles overlap in space. Constraint $T$ checks whether CNOTs overlap in time.
For any pair of CNOTs $g_i$ and $g_j$, they cannot overlap in time if they overlap in space: $
S(g_i,g_j) \implies \lnot T(g_i,g_j)$.

{\noindent \bf One Bend Paths: }
In this policy, CNOT routes are restricted to the two paths along the bounding rectangle of the control and target qubit. For example, in Figure \ref{figOBP}, the CNOT is allowed to use one of the two highlighted paths. To implement this policy, the solver selects one of the two routes for every CNOT in the program.

To express constraints for this policy, we use variables to record the junction through which the CNOT is routed. The one bend path is composed of two segments: control to junction and junction to target. For generality, we can consider these segments as rectangles, and apply the same overlap check as in rectangle reservation. Denote the control to junction path for CNOT $i$ as $R_i^{cj}$. Then, we can check if two CNOTs $g_i$ and $g_j$ overlap using:
\begin{align}
Overlap(i,j)= &  S(R^{cj}_i, R^{cj}_j)  \lor S(R^{cj}_i, R^{jt}_j)  \lor \notag \\
& S(R^{jt}_i, R^{cj}_j)  \lor S(R^{jt}_i, R^{jt}_j) 
\end{align}
Similar to rectangle reservation, we impose the condition that CNOTs do not overlap in time if they overlap in space.

\subsection{Reliability Constraints}\label{sec:52}
To optimize the reliability of program executions, we use a set of constraints to track the reliability scores of CNOT and readout operations in the program. Let $g.\epsilon$ denote the reliability score for the operation $g$. For readout operations, we set the reliability as 
\begin{align}
    \forall g \in G_{Readout}: \forall h \in Q_H: g.q = h \Rightarrow g.\epsilon = E^{R}_h
\end{align}
where $E^{R}_h$ is the reliability score for readout operations on hardware qubit $h$, and $G_R \subseteq G$ is the set of readout operations.

In \rSMTstar\, we perform reliability optimization using the one bend paths routing policy. Under this policy, for CNOT gate, we set reliability tracking variables based on the junction used for routing.  For each pair of hardware qubits, we compute the reliability of the two possible paths, and store them in a matrix $E^{C}$, indexed by the hardware qubits and junction. This reliability factors in the reliability of the swap paths through the junction and the actual CNOT operation. Let $g.j$ be the junction for gate $g \in G_{CNOT}$. The constraints to track CNOT error are given for all $g \in G_{CNOT}$ and for all $h_1, h_2, h_j \in Q_H$:
\begin{align}
    g_c = h_1 \land g_t = h_2 \land g.j = h_j \Rightarrow g.\epsilon = E^C_{h_1, h_2, j} 
\end{align}
In our experiments, considering the error rates of single qubit gates such as H, X, Y etc. is not required for \ibmnameshort, because their error rates are much smaller than CNOTs and readouts. For systems where such errors matter, they can be easily incorporated into the optimization using similar constraints.

\subsection{Optimal Compilation: Objective Function}\label{sec:objective_function}

The different optimization variants use different objective functions.  For the time-oriented variants \tSMT and \tSMTstar, the objective function is based on the execution time for the program. Using the gate scheduling and duration constraints in Section \ref{sec:constraints}, the objective is to minimize the finish time of the last gate in the dependency order.

For the reliability-oriented variant, \rSMTstar, the objective function is based on the reliability of a program execution. 
We define the reliability of a program execution as the product of the reliability of each of its gates. Since single qubit gates have low error, we define the reliability using CNOT and readout operations only. Ideally, the reliability objective would be the product across all gates of the readout and CNOT errors for the whole program: $ \max \prod_{\forall g \in G_{Readout} \cup G_{CNOT}} (g.\epsilon) $.  Because the SMT solver requires linear operations, we convert this to an additive linear objective function by considering the logarithm of the operation reliabilities, instead of their product.  Finally, to allow for different emphases on readout error versus CNOT error, we convert the above objective into a weighted objective using a weight $\omega$ which is applied to the readout error rates:
\begin{align}\omega\sum_{g \in G_{Readout}}\log (g.\epsilon) + (1-\omega)\sum_{g \in G_{CNOT}} \log (g.\epsilon).\end{align} 
We use this objective to study the relative importance of CNOT and readout error rates. 

Optimizing reliability places qubits at hardware locations with high CNOT and readout reliability. It indirectly optimizes qubit movement because CNOT gates between non-adjacent qubits have low reliability. This objective is used by \rSMTstar\ in our experiments. The output of the solver has the optimal reliability with respect to the program and machine model assumptions. Our experiments show that it is also near-optimal in execution duration. 

To compute a qubit mapping and gate schedule which maximizes this objective, we set up an optimization problem using this along with the mapping and scheduling constraints, gate durations using calibration data, routing approaches, and reliability constraints discussed before. The reliability constraints make the $g.\epsilon$ variables dependent on the qubit mapping variables. 



\section{Heuristic Compilation}\label{sec:heuristic}
Where tractable, the SMT-based compilation approach offers the best chance at successful application runs on real hardware. However, effective heuristic approaches may offer similar reliability but scale better to future NISQ systems with hundreds of qubits. Here we propose and evaluate heuristic mapping/scheduling alternatives as comparators to the optimization-based approaches. 

Our heuristic techniques are also based on a program graph constructed from the program IR. The program graph has a node for every qubit, and an edge between every pair of qubits which is involved in a CNOT. For example, the program graph of BV4 has 4 nodes for $p_{0,1,2,3}$ and 3 edges, one from each of $p_{0,1,2}$ to $p_3$. 
For each heuristic, we first compute the most reliable path between every pair of hardware qubits using Dijkstra's algorithm, where edge weights are given as the negative log of the CNOT errors from the calibration data. For both heuristics, once we map the qubits, we schedule gates using an earliest ready gate first policy \cite{heckey1} and route based on the precomputed paths.

\subsection{Greatest Vertex Degree First}
The \greedyV\ heuristic seeks to minimize communication distance (and therefore reduce the number of error-prone SWAP operations) by considering qubits in descending order of degree. The degree of the qubit is the number of CNOTs in which the qubit is used. First, place the highest degree program qubit at the hardware location which has highest readout reliability among high degree hardware qubits. Next, for each program qubit which shares a CNOT with an already placed qubit, place this qubit in order to maximize the total reliability of paths between it and each of its placed neighbors, where the total reliability is given by the sum of the path lengths computed between it and its neighbors.

\subsection{Greatest Weighted Edge First}
In \greedyE, we map edges in the descending order of weight. The weight of an edge between two nodes is the number of times a CNOT gate is invoked between them. Therefore, placing edges with high weight first allows qubits which interact highly to be close together. Such placement reduces qubit movement and increases reliability. The algorithm starts by placing the highest weighted edge at on hardware location with maximum CNOT and readout reliability.
Next, for each edge which has one mapped one unmapped endpoint, we map the unmapped qubit to the position which maximizes the total reliability of CNOTs with already mapped qubits, where the total reliability is given by the sum of the path lengths computed from before between it and its neighbors. The process is repeated for each unmapped edge in weight order.

%% file: txt/expt.tex
\section{Experimental Setup}\label{sec:expt_setup}
\paragraph{Benchmarks:}
Table \ref{tab:benchmarks} lists 12 quantum programs derived from prior work on compilation and system benchmarking \cite{bench1, bench2, bench3}. These benchmarks include the Bernstein-Vazirani algorithm \cite{bernsteinvazirani}, Hidden Shift Algorithm \cite{hiddenshift}, Quantum Fourier Transform \cite{NielsenChuang}, a one bit adder and important quantum kernels such as the Toffoli gate \cite{Mermin}. We used or created Scaffold programs for each benchmark and obtained LLVM IR using the ScaffCC compiler \cite{scaffcc1}. To be runnable on real-system QC hardware, the benchmarks must be relatively small in qubit counts and short in execution time steps.  Nonetheless, our ability to show order-of-magnitude improvements in success rate for these programs is a promising indicator of the value of such compilation techniques for future larger systems and programs.  Furthermore, several of these programs---such as QFT and Toffoli---are important kernels for larger programs.  

Beyond these, to study scalability trends across different qubit and gate counts, we generate a synthetic benchmark where we can specify the number of qubits and gates and from this, we experiment with randomly generated quantum programs with $4$-$128$ qubits and $128$-$2048$ gates. We generate these circuits by uniformly sampling gates from the universal gate set of {H, X, Y, Z, S, T, CNOT}.
\begin{table}[t]
\centering
\includegraphics[scale=0.4]{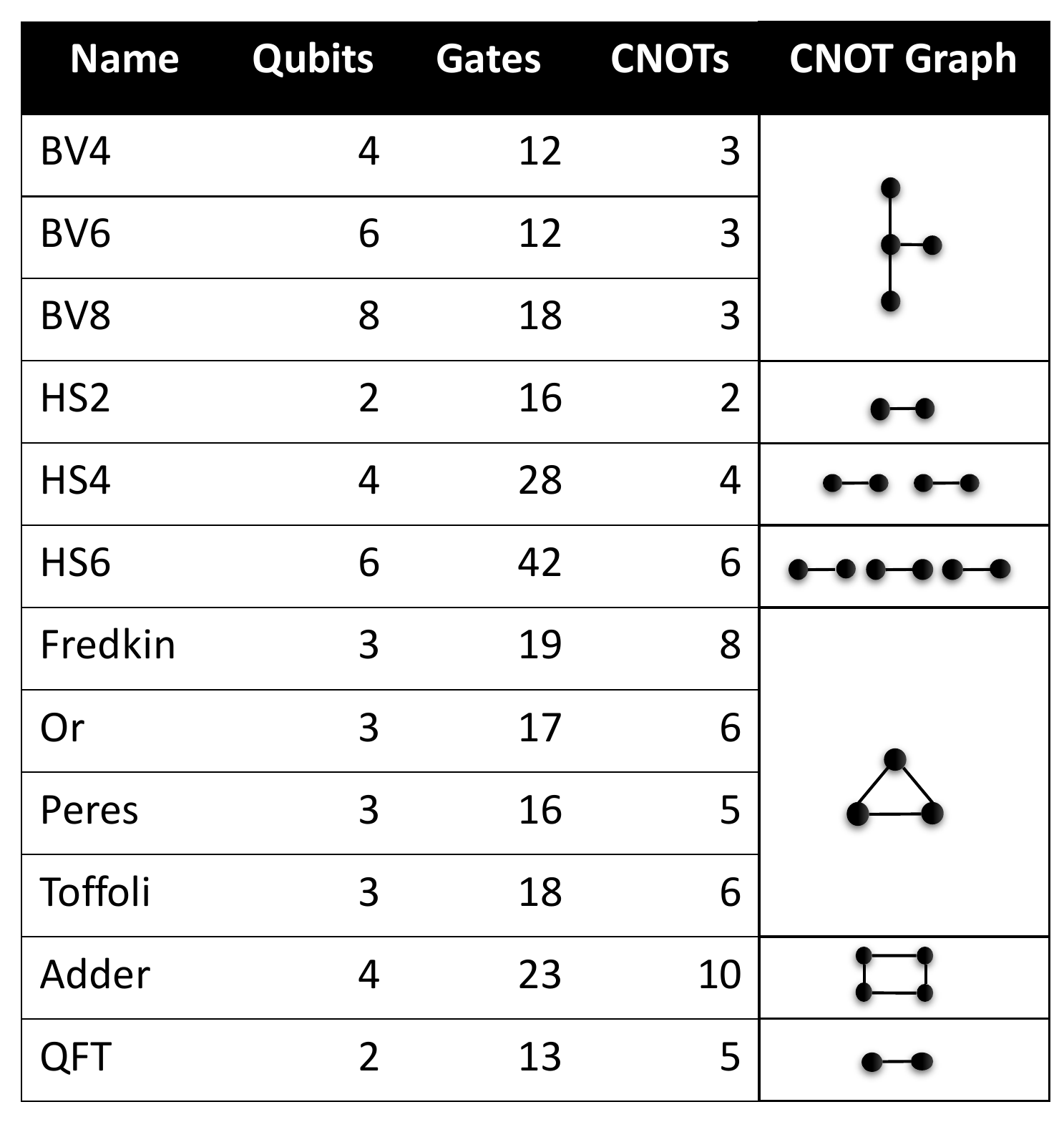}
\caption{Characteristics of benchmark programs.}
\label{tab:benchmarks}
\end{table}

\paragraph{Compiler Configurations:}
To study various compilation schemes, our framework includes various options for the solver, routing policy, use of calibration data and other parameters. We evaluate these options one factor at a time using the configurations listed in Table \ref{tab:compiler_config}. We compare \rSMTstar and \tSMTstar to demonstrate the benefits of noise-adaptive compilation. We compare \tSMTstar and \tSMT\ to demonstrate the importance of considering gate times and coherence times from calibration data. 

\paragraph{Experimental Setup:}
Our compilation experiments use an Intel Skylake processor (2.6GHz, 12GB RAM) using Python3.5 and gcc version 5.4. Our optimization approach uses the Z3 SMT solver \cite{z3}.  To perform experiments on \ibmnameshort, we use the IBM Quantum Experience APIs \cite{ibmq, ibmqexp}. The daily machine calibration data is available through the Quantum Experience APIs. The calibration data includes time data such as single qubit gate time, qubit coherence time (T2 time), durations for CNOT gates, and error rates such as single qubit gate error, CNOT gate error, and read out (measurement) error.  We use IBM's Qiskit compiler/mapper as our baseline for comparison, version 0.5.7. 

\paragraph{Metrics:} Before each run, we obtain the latest calibration data, and recompile the benchmark. We execute each benchmark on \ibmnameshort, using 8192 trials in each run. We measure the success rate as the fraction of trials which gave the correct answer. For example, success rate of $0.6$ means the execution produced the correct answer in $60\%$ of the trials. The ideal success rate is $1$, where all trials succeed. Results within a single graph are performed closely in time so are comparable. Results from different graphs may not be comparable because the machine error characteristics can be different across runs. We also study quantum execution time and compilation time. Because timing granularity is so coarse, execution time is estimated using real gate duration data from the \ibmnameshort\  system. We report durations in terms of timeslots on \ibmnameshort, where each timeslot is 80ns.

%% file: txt/results.tex
\section{Optimizing Execution Reliability}\label{sec:opt_reliability}

\begin{figure}[t]
    \centering
    \includegraphics[scale=0.3]{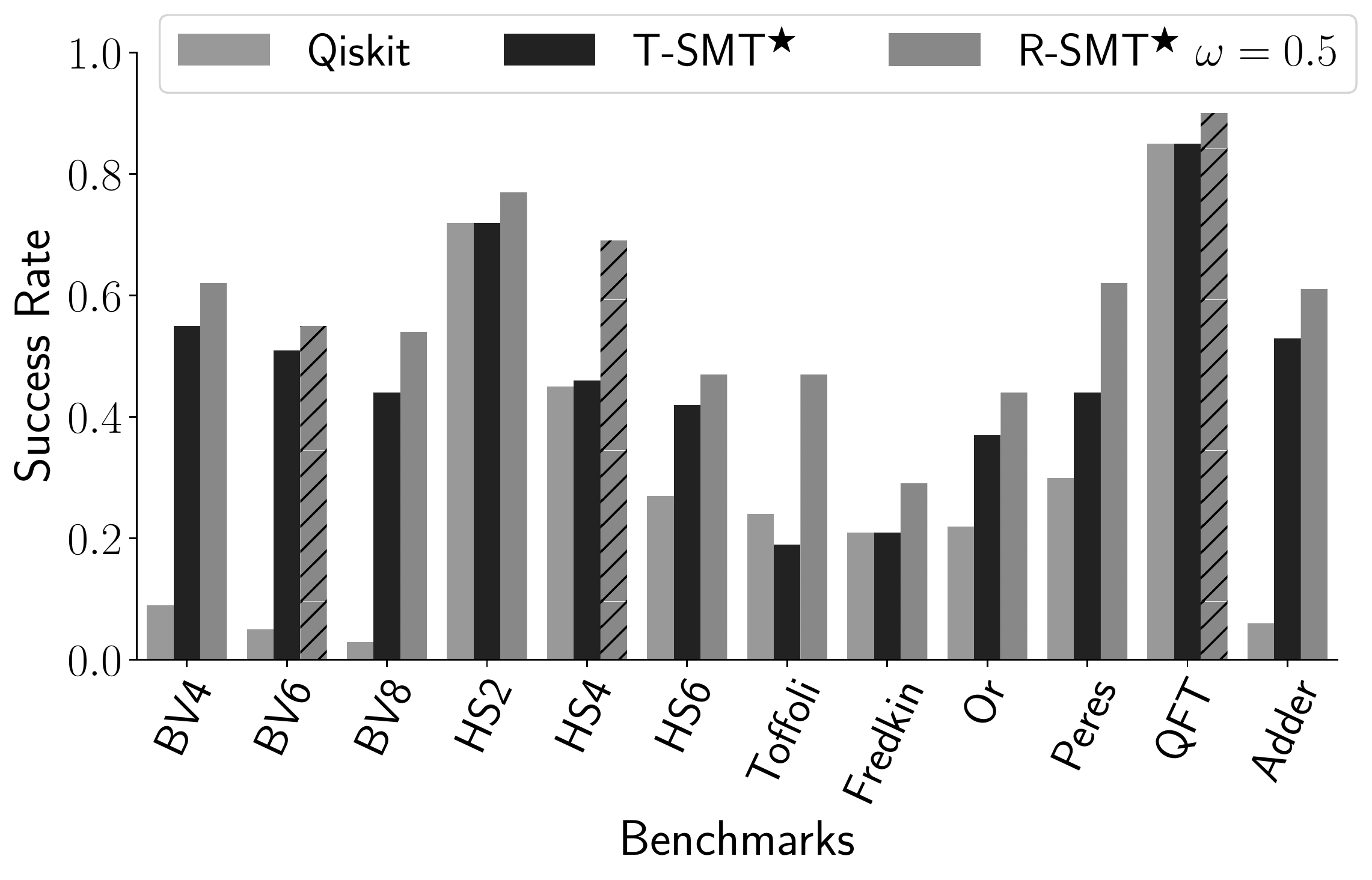}
    \caption{Measured success rate of \rSMTstar compared to Qiskit and \tSMTstar. (Of 8192 trials per execution, success rate is the percentage that achieve the correct answer in real-system execution.)  \rSMTstar obtains higher success rate than Qiskit because it simultaneously adapts placement according to dynamic error rates and avoids unnecessary qubit movement.}
    \label{fig:err_full}
\end{figure}

\begin{figure}[t]
    \centering
    \includegraphics[scale=0.5]{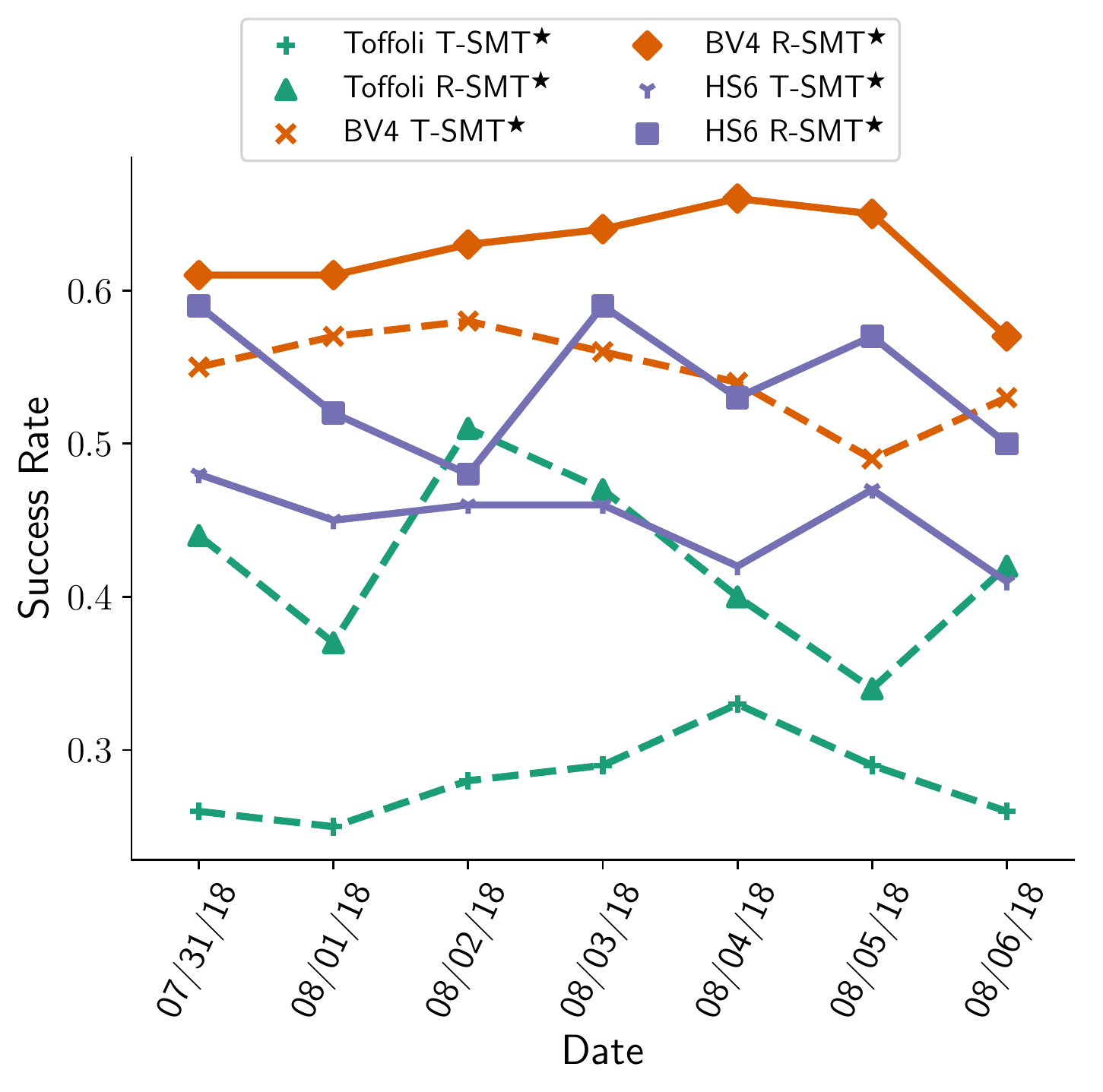}
    \caption{Executions of three benchmarks for 1 week. \rSMTstar is more resilient to errors compared to \tSMTstar. Similar trends for other benchmarks.}
    \label{fig:daily_toffoli}
\end{figure}

\begin{figure*}[t]
\centering

\subfloat[Success Rate]
{
    \includegraphics[scale=.22]{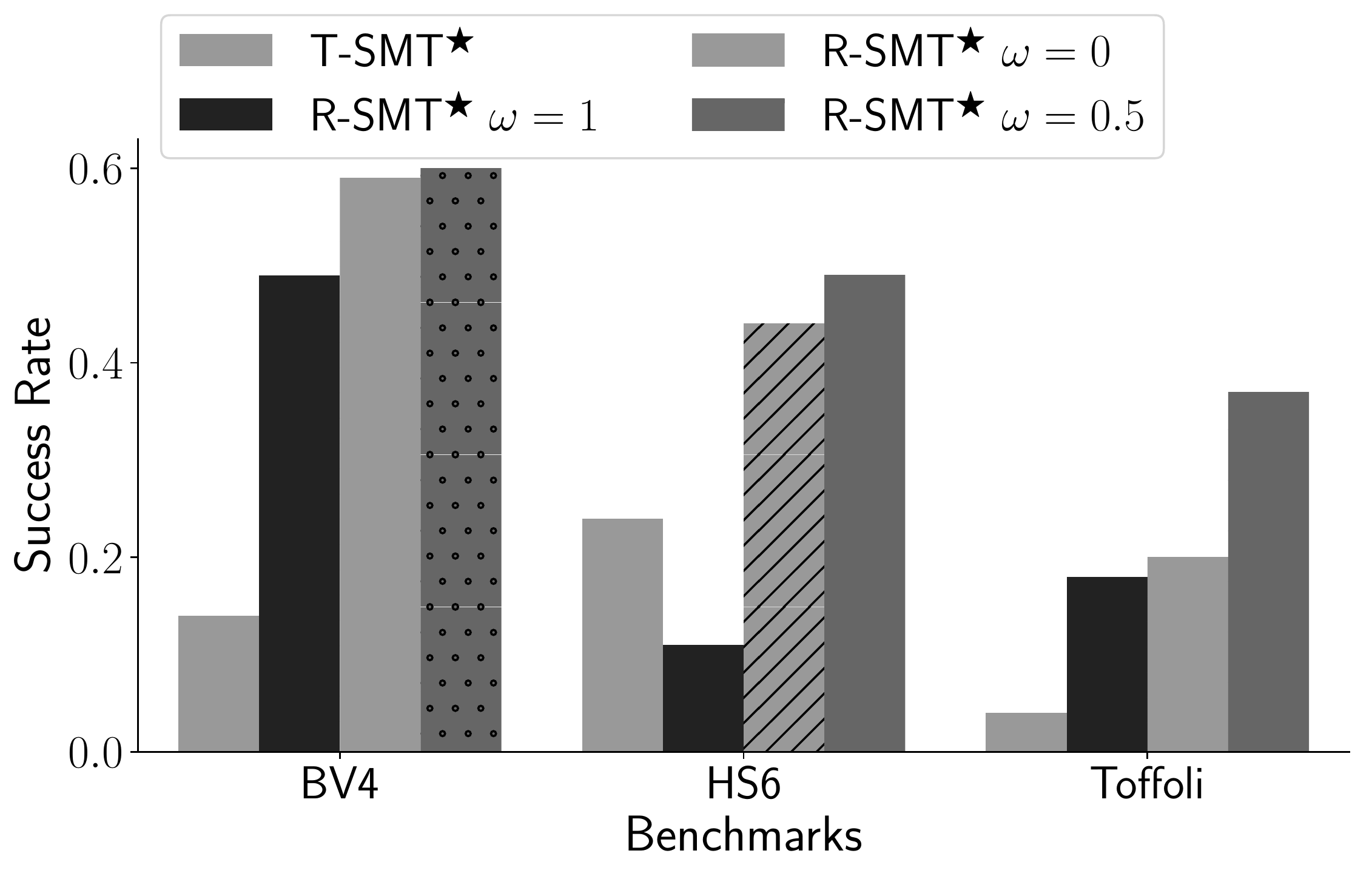}
    \label{fig3}
}
\quad
\subfloat[Execution Duration]
{
    \includegraphics[scale=.22]{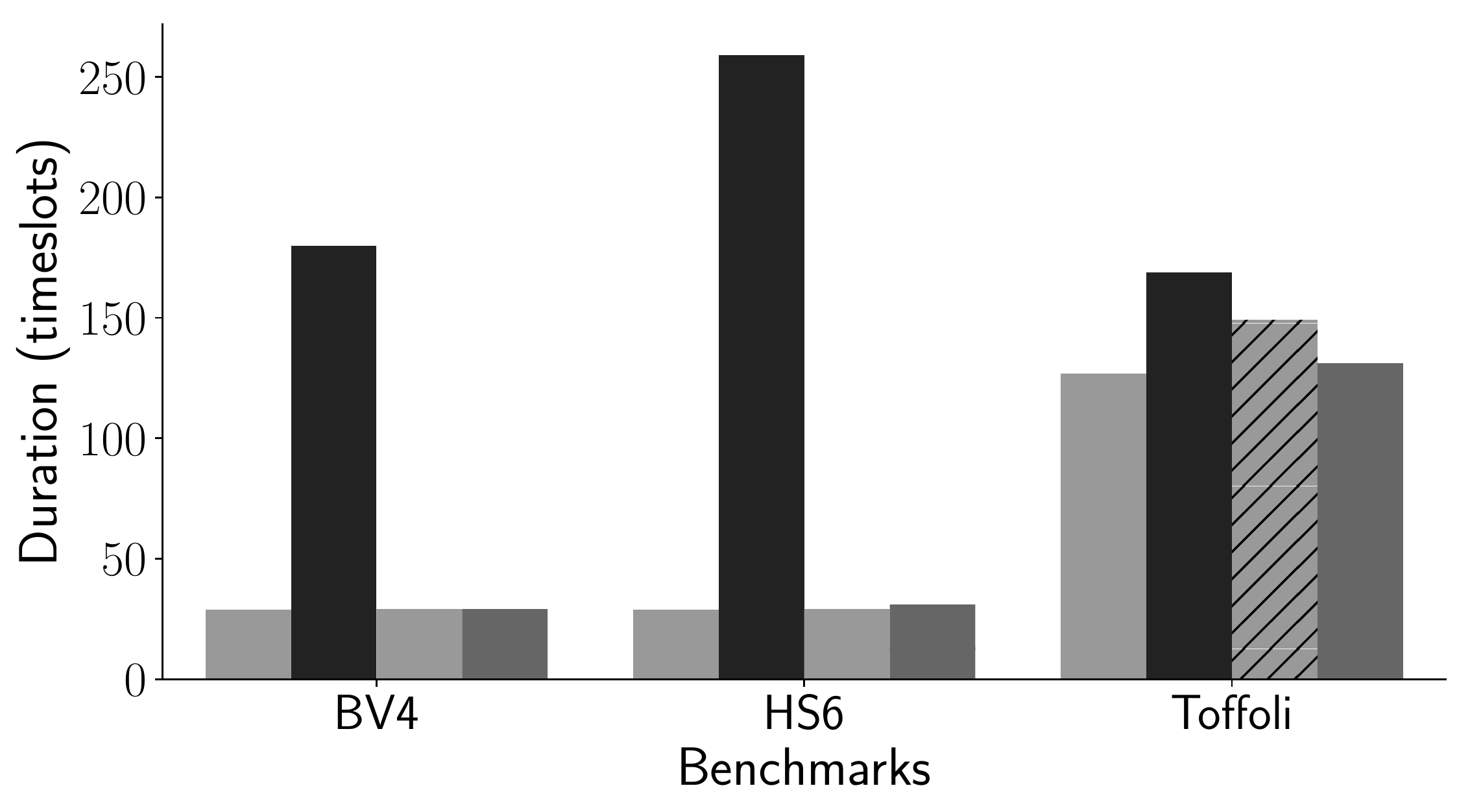}
    \label{fig2}
}
\quad
\subfloat[Compile Time]
{
    \includegraphics[scale=.22]{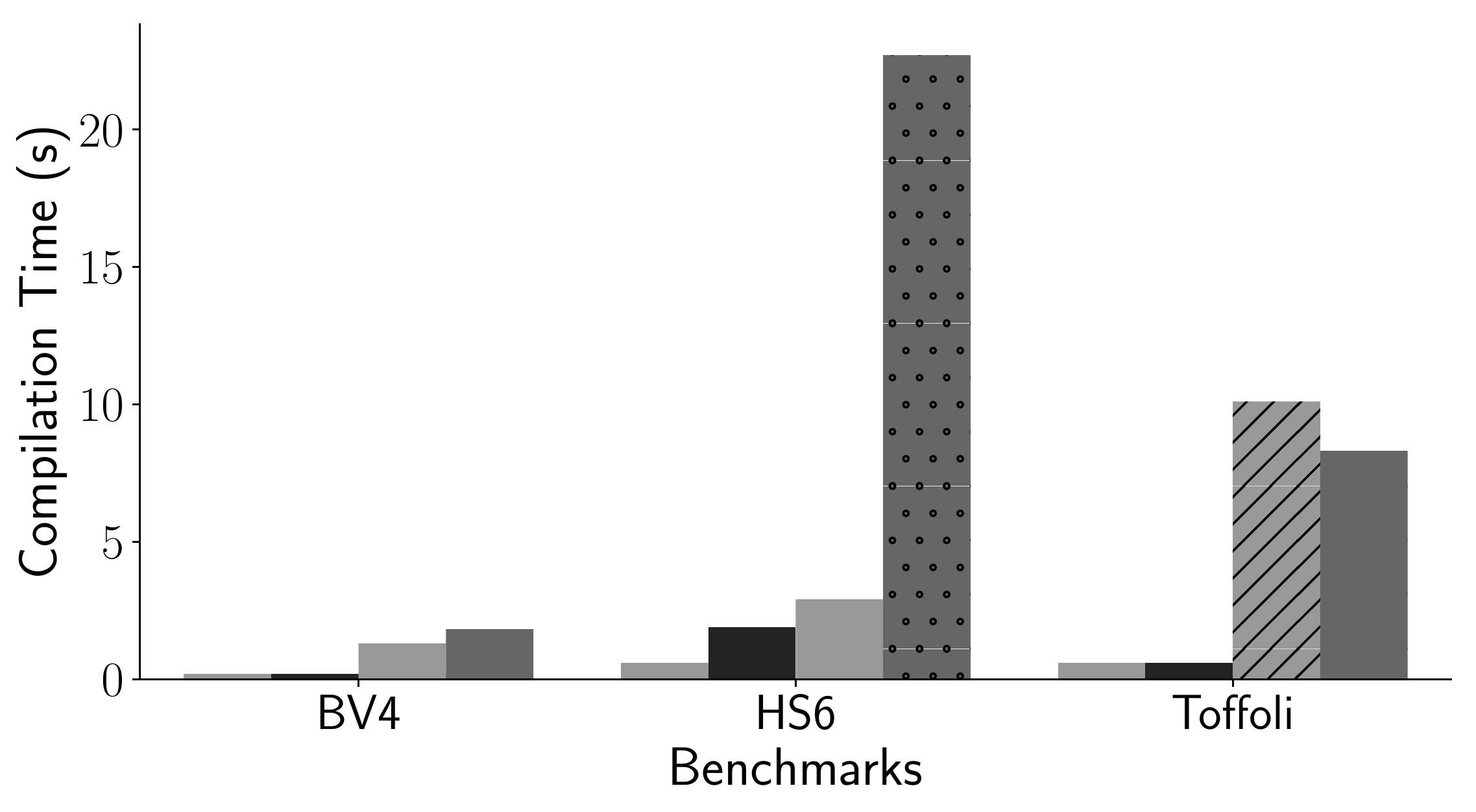}
    \label{fig}
}
\caption{Measured success rate, execution duration and compile time for three representative benchmarks. \tSMTstar\ which directly optimizes for execution duration obtains the minimum execution durations, but \rSMTstar\ with $\omega=0.5$ is close, and more resilient to errors (higher success rate). All benchmarks compile in less than 1 minute.}
\label{fig:res_3bench}
\end{figure*}
\begin{figure*}[t]
   \centering
\subfloat[\Qiskit]
{
    \includegraphics[scale=.6]{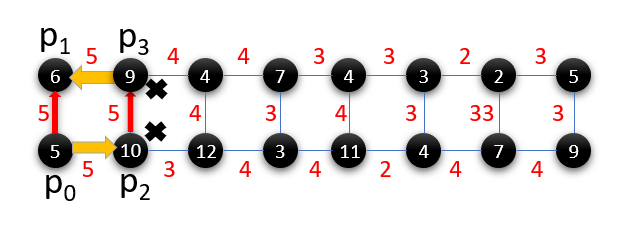}
}
\subfloat[\tSMTstar:Optimize duration without error data]
{
    \includegraphics[scale=.6]{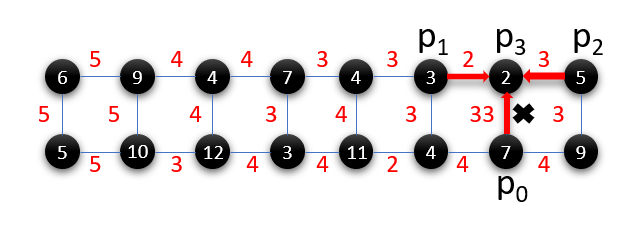}
}\vspace{-4mm}
\subfloat[\rSMTstar ($\omega=1$): Optimize readout reliability]
{
    \includegraphics[scale=.6]{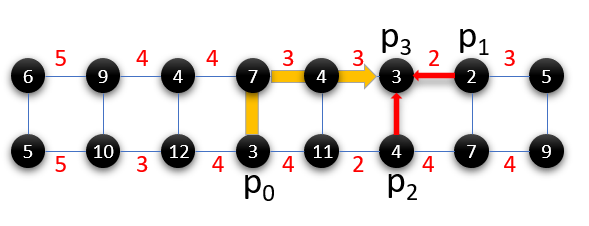}
}
\subfloat[\rSMTstar ($\omega=0.5$): Optimize CNOT$+$readout reliability]
{
    \includegraphics[scale=.6]{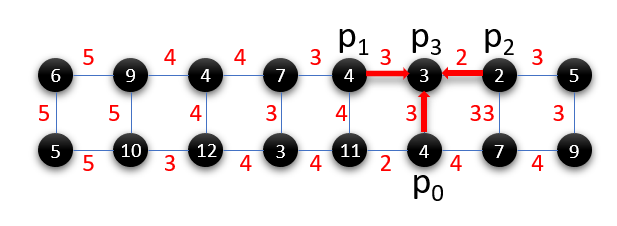}
}
   \caption{For real data/experiment, on \ibmnameshort, qubit mappings for Qiskit and our compiler with three optimization objectives, varying the type of noise-awareness. In each figure, the edge labels indicate the CNOT gate error rate ($\cross10^{-2}$), and the numbers inside each node indicate that qubit's readout error rate ($\cross10^{-2}$). The thin red arrows indicate CNOT gates. The yellow thick arrows indicate SWAP operations. (a) \Qiskit\ finds a mapping which requires SWAP operations and uses hardware qubits with high readout errors  (b), \tSMTstar finds a a mapping which requires no SWAP operations, but it uses an unreliable hardware CNOT between $p_3$ and $p_0$. (c) Program qubits are placed on the best readout qubits, but $p_0$ and $p_3$ communicate using swaps. (d) \rSMTstar finds a mapping which has the best reliability where the best CNOTs and readout qubits are used. It also requires no SWAP operations.}
   \label{fig:bv_all_mappings}
\end{figure*}
\begin{figure}[t]
    \centering
    \includegraphics[scale=0.3]{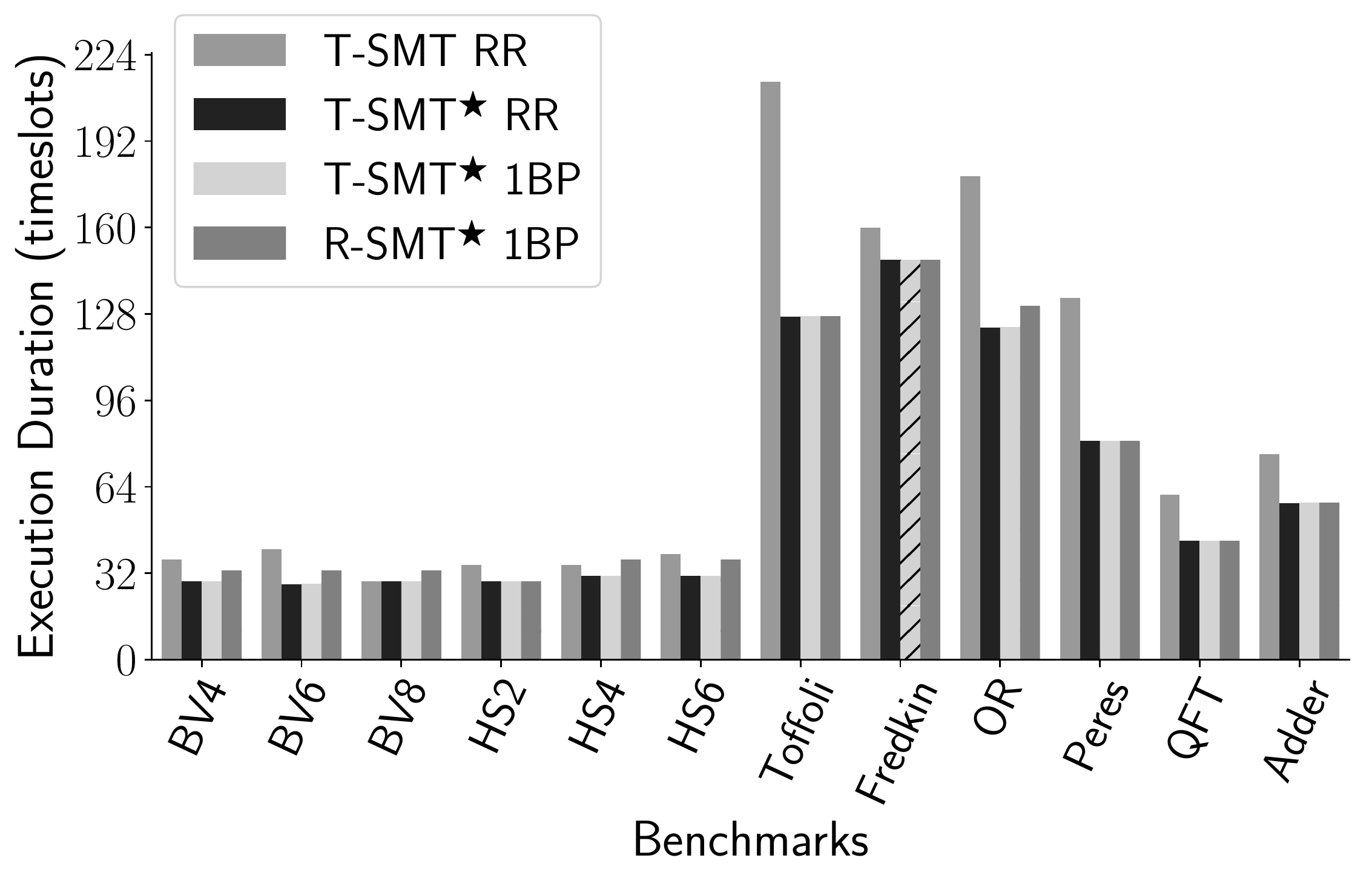}
    \caption{Effect of gate durations, routing policy and objective function on execution duration. Although reliability is our primary objective, several variants perform well on run time as well. \tSMTstar (either RR or 1BP) has the best execution duration, but \rSMTstar is very close in run time and offers better success rates.  Noise-aware policies, \rSMTstar and \tSMTstar, are  $1.6$x better than \tSMT.  }
    \label{fig:time_params}
\end{figure}
\begin{figure}[t]
    \centering
    \includegraphics[scale=0.3]{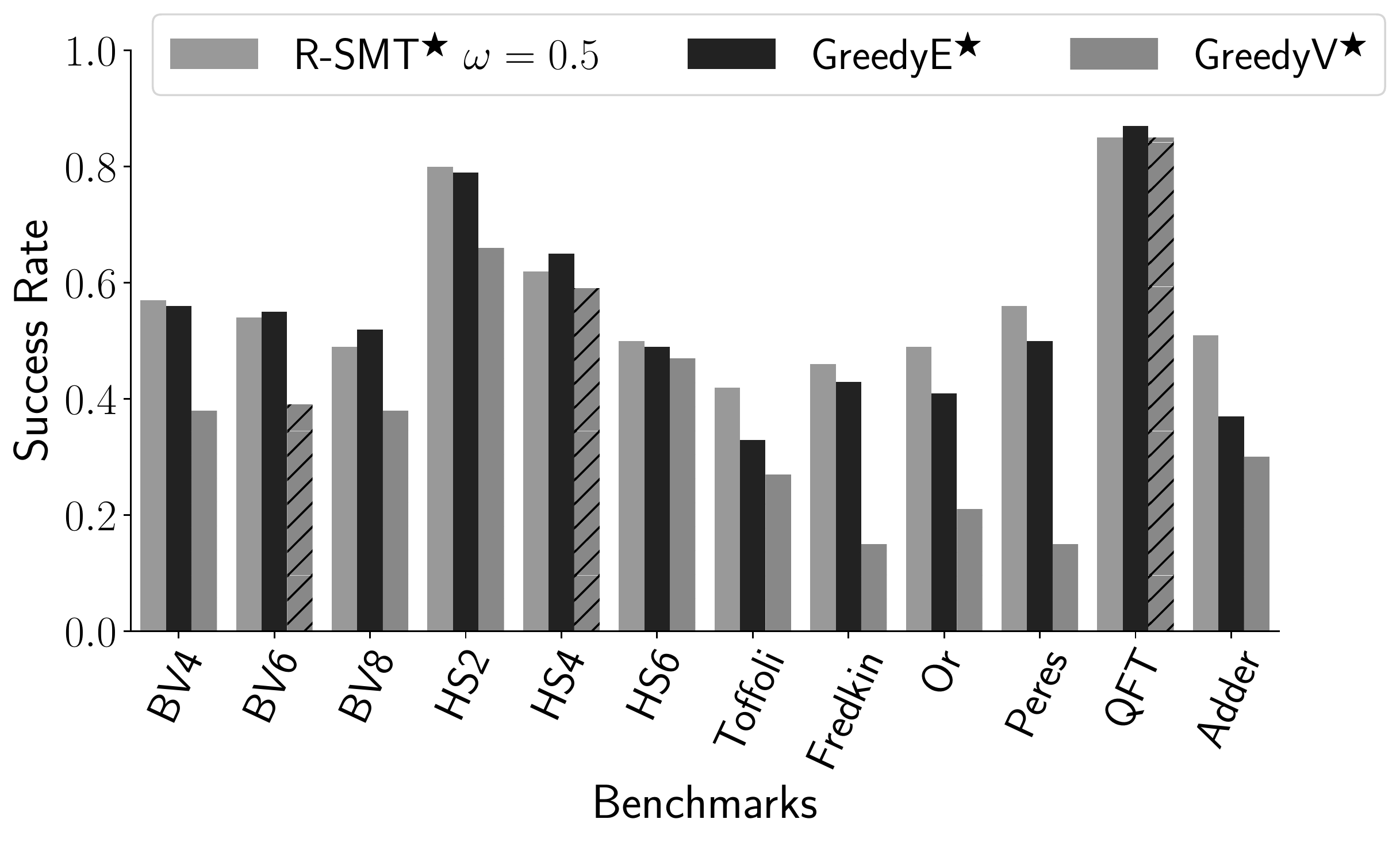}
    \caption{Noise-aware Heuristics: \greedyE\ heuristic mapping offers reliability comparable to \rSMTstar on most benchmarks.}
    \label{fig:greedy_results}
\end{figure}

\par{\textbf{Baseline Comparison to IBM Qiskit}:} We compare the success rate of program runs from our compiler versus the IBM Qiskit compiler for real-system runs on \ibmnameshort. Figure \ref{fig:err_full} shows the success rate  of the IBM Qiskit compiler, \tSMTstar\ and \rSMTstar\ with $\omega=0.5$ on all the benchmarks. In all benchmarks, \rSMTstar\ has higher success rate than Qiskit, indicating that its reliability-oriented objective function is effective. In fact, \rSMTstar\ obtains geomean $2.9$x improvement over Qiskit, with up to $18$x gain. Figure \ref{fig:bv_all_mappings} shows the mapping used by Qiskit,  \tSMTstar and \rSMTstar for BV4. Qiskit places qubits in a lexicographic order without considering CNOT and readout errors and incurs extra swap operations. For BV8, the compiled code produced by Qiskit used 15 CNOT operations to move qubits (in addition to the 3 CNOTs required by the algorithm), while \rSMTstar\ obtains a mapping which require no qubit movement. Each extra CNOT gate increases both the error rate and the execution duration of the code and leads to poor success rate. Benchmarks which require no qubit movement such as BV, HS, QFT and Adder have higher reliability than Toffoli, Fredkin, Or, and Peres, which require at least one qubit swap.

In all benchmarks, \rSMTstar\ outperforms \tSMTstar, even though they use the same number of qubit movement operations. While optimizing qubit communication is important, it is essential to optimize for gate error rates to improve success rate. In fact, in our experiments, when the machine state has high variability, \rSMTstar\ can obtain up to $9.2$x improvement in success rate over \tSMTstar\ (see Fig.  \ref{fig:res_3bench} and \ref{fig:bv_all_mappings}).

\par{\textbf{Resilience to Daily Variations}:} Since IBM limits the executions researchers may perform per day, we perform detailed experiments on three benchmarks, BV4, HS6 and Toffoli. These benchmarks are chosen as examples of different CNOT patterns (see Table \ref{tab:benchmarks}). Figure \ref{fig:daily_toffoli} compares the success rate of \rSMTstar\ and \tSMTstar\ over a week for the three benchmarks. The success rate of the programs change every day because error rates of the hardware CNOT and readout units change daily. (We recompile each day before running.) For all three benchmarks, \rSMTstar\ is more resilient to error than \tSMTstar, since it adapts the qubit mappings to account for daily variations in operation error rates. Since \tSMTstar\ compiles based on static information (qubit topology and gate duration), it uses the same qubits and hardware gates every day, irrespective of their dynamic error characteristics.
\subsection{Choice of Optimization Objective}  
Figure \ref{fig:res_3bench} compares \rSMTstar\ with $\omega=\{0, 0.5, 1\}$ and \tSMTstar\ on the three benchmarks. \rSMTstar\ with $\omega=0.5$ achieves the highest success rate among the methods, with up to $9.25$x gain over \tSMTstar. 
For BV4, we illustrate the mappings obtained by the these methods in Figure \ref{fig:bv_all_mappings}. \tSMTstar\ obtains a mapping which requires no qubit movement, but it uses a hardware CNOT with very high error rate. With $\omega=1$, \rSMTstar\ optimizes only for readouts and uses long swap paths which reduce success rate. With $\omega=0.5$, \rSMTstar\ maps qubits to simultaneously optimize CNOT gate error, readout error and qubit movement.

\rSMTstar\ with $\omega=0.5$ also achieves near-optimal execution durations, comparable to \tSMTstar, which directly optimizes for duration. From the perspective of compilation time, optimizing for reliability is harder than optimizing execution duration. However, each method finds optimal mappings in under a minute, for each benchmarks. 

\rSMTstar\ was executed with $\omega \in [0,1]$ to determine the relative importance of optimizing for readout error and CNOT error. In general, choosing an $\omega$ roughly near $0.5$ is appropriate to obtain good success rates. On the \ibmnameshort\ machine, readout and CNOT error rates are fairly balanced, and hence we see that an equal weighted combination of both is suitable for optimization.

\subsection{Sensitivity to Gate Durations and Coherence Time}

We test whether the use of real gate time data significantly affects the execution duration of NISQ benchmarks. Our compiler is run on three settings: \tSMT (RR) which assumes all hardware CNOTs have the same gate duration and \tSMTstar\ (RR) and \rSMTstar\ (1BP) which use real gate durations. We restrict \rSMTstar\ to the 1BP policy to reduce the number of experimental configurations; we show in Section \ref{sec:routing_policy} that the choice of routing policy doesn't affect execution duration for NISQ benchmarks.

\par{\textbf{Gate Durations}:} Figure \ref{fig:time_params} shows execution duration, computed using the gate time data, for the three methods. Considering real gate durations can improve the execution duration for each benchmark, with up to $1.68$x gain on Toffoli. Considering real durations increases the number of constraints in the optimization problem and increases the compilation time by up to $3$x (not shown). Even with real durations, each benchmark requires only a few seconds of compilation time. 

\par{\textbf{Coherence Time}:} Each benchmark finishes in less than 150 timeslots using the \rSMTstar\ method. Since the coherence time of the worst qubit on the machine is more than 300 timeslots, considering fine grained variations in coherence time is not necessary for our benchmarks. 

\subsection{Effect of Routing Policy}
\label{sec:routing_policy}
Figure \ref{fig:time_params} compares the execution duration and compilation time of \tSMTstar\ with two routing policies (RR and 1BP) and \rSMTstar\ (1BP). The three policies produce executables with similar execution duration since NISQ benchmarks are small, and have only few parallel CNOTs. Hence, most CNOTs execute without swapping or blocking qubits. Although \rSMTstar\ optimizes reliability, it obtains execution durations close to \tSMTstar\ on all benchmarks.


\subsection{Success Rate and Scalability of Heuristics}\label{sec:scalability}
We compare the success rate of heuristics to the optimal methods and evaluate the scalability of all methods.

Figure \ref{fig:greedy_results} compares the success rate of the heuristics and \rSMTstar. Greedy methods are comparable to \rSMTstar\ in success rate and in some cases, they outperform \rSMTstar\ marginally because $\omega=0.5$ may not the optimal value for every benchmark and machine state. \greedyE\ is as successful as \rSMTstar\ in all cases. Our study reveals the edge based heuristic \greedyE, is more successful than the vertex based heuristic \greedyV. Considering edges instead of vertices allows the heuristic to prioritize the reliability of the most frequent CNOTs.

\begin{figure}[t]
    \centering
    \includegraphics[scale=0.4]{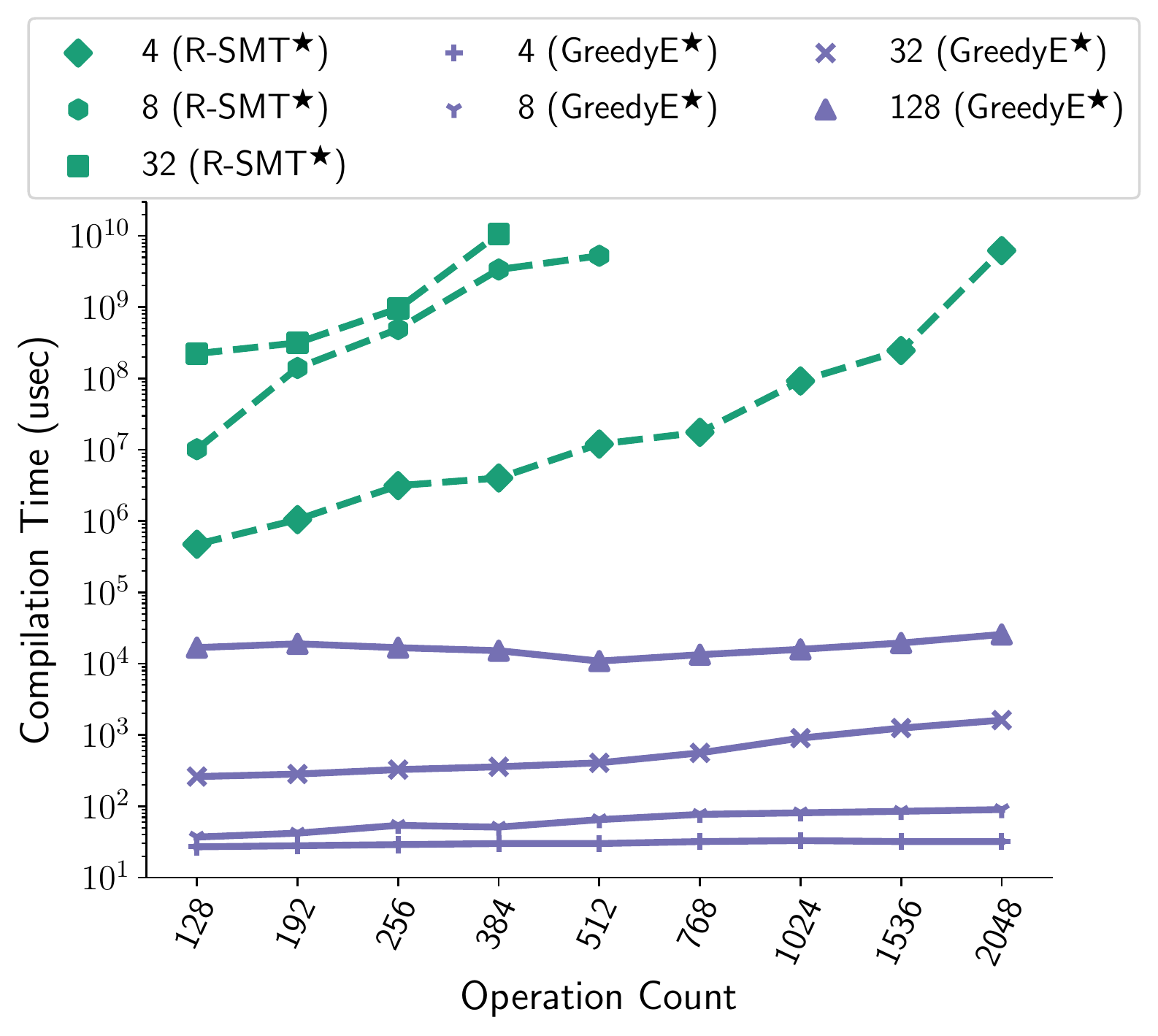}
    \caption{Scalability of optimal and heuristic methods on synthetic benchmarks. The legend shows a line's qubit count.}
    \label{fig:scale_study}
\end{figure}

To study the scalability of optimal and heuristic methods, we used a benchmark of randomly generated quantum programs. Figure \ref{fig:scale_study} shows the compilation time on the benchmark. \rSMTstar\ requires up to 3 hours to compile a program with 32 qubits and 384 gates. On the other hand, the greedy methods compile programs in under one second in all cases.

%% file: txt/related.tex
\section{Related Work}\label{sec:related_work}
\label{sec:related}
Quantum programming languages and their compilers have been developed by extending languages such as C and C$\#$ with quantum functionality. Examples include Quipper \cite{quipper1, quipper2}, LIQUi$\ket{}$ \cite{liquid1}, and Scaffold \cite{scaffcc1, scaffcc2}. ProjectQ \cite{projectq1} is a Python framework to describe quantum circuits and compile them for different backends. PyQuil, developed at Rigetti \cite{pyquil, forest} is another such Python framework. Until very recently, most backends were simulators or resource-estimators, rather than real hardware. Our work here is an early example of top-to-bottom compilation from a high-level QC language (Scaffold) to real hardware. 

OpenQASM \cite{openqasm1} and Quil \cite{quil} are low-level assembly language interfaces to QC hardware \cite{ibmq}. To target IBM machines, our compiler produces optimized OpenQASM code. Our compiler can be easily extended to generate code for other low-level interface languages also.

QC compilation has been studied for different hardware technologies and topologies. \cite{intel1} develops a heuristic to schedule quantum circuits on linear topologies where all gates (including swaps) consume unit time. \cite{ai1} uses AI planners for scheduling a specific class of quantum circuits. \cite{heckey1} develops heuristic techniques for ion trap systems. Recently, \cite{cgo18} compiled small benchmarks for IBM systems, based on only qubit topology information, not calibration data. Two recent works \cite{qx21, qx22}  reduce swap operations and optimize 1-qubit gates for 5-qubit IBM systems. Other prior work \cite{sched1, sched2, sched3, sched4, sched5, sched6, sched7, sched8, sched9, sched10, sched11, sched12, sched13, sched14} are either manual methods or restricted to a specific architecture, or a specific class of quantum programs; none account for  real gate durations, gate errors and variations in qubit coherence time. Similarly, other work has focused on compilation issues in future QC systems with ECC \cite{future1, future2, future3, future4, future5, future6, future7}. In contrast to these works, our compiler is designed and evaluated using a real IBM QC system. Using real-system measurements, we show that driving compilation decisions based on machine calibration and configuration data dramatically improves program success rates.

\cite{calib1, tannu_qureshi} observed the usefulness of calibration data. While \cite{calib1} uses error data manually to improve execution success, \cite{tannu_qureshi} proposes the use of calibration data-aware qubit mapping and movement policies on the 20-qubit IBM system. However, they do not perform any real hardware executions of their mapped code, making it difficult to compare results based on reliability. Their work also does not discuss how program success rates are computed on the simulator and uses error rates which are scaled by $10$x. Simulated or scaled success rates may not correlate well with real performance. \cite{hardvardallocator} is another recent work which maps circuits in described in the low-level OpenQASM language to \ibmnameshort. Their simulated annealing based method considers only CNOT error rates to compute the qubit mapping. In contrast, our work develops a toolflow which maps high-level programs onto \ibmnameshort, using both CNOT and readout error rates, gate times, coherence times and qubit layout. Using real-system  evaluations our work determines the relative importance of these parameters and compares the performance of heuristic and optimal techniques.

\nocite{koen}

%% file: txt/conclusions.tex
\section{Conclusions}\label{sec:conclusions}
This paper proposed and evaluated calibration-aware compiler techniques for NISQ systems. We considered optimal and heuristic compilation methods, the use of calibration data, different objective functions and routing policies. Our evaluations show it is crucial to adapt quantum program compilation to dynamic operation error characteristics of the machine. It is most important to consider CNOT and readout error rates, since these operations are more noisy than single qubit gates. Optimization based on qubit coherence time is also useful, but less critical here because gate errors severely limit useful computation time. Our research has shown that SMT approaches are very effective for current and near-term systems, but may not scale well to the far-NISQ machines of 500 qubits or more. For those, we have developed heuristic approaches, \greedyV\ and \greedyE, which offer nearly as good results but with much more tractable compile times.

This paper's results offer important insights on QC based on real-system measurements. Our work shows the importance of initial qubit placement. Namely, benchmarks which require more qubit movement are hard to reliably execute on systems with grid topologies. Our results show that proper placement could result in over 10X improvements in run success rate. Mapping and scheduling based on calibration data offer further benefits. Ultimately the best-performing approach offered up to $18$X improvement ($2.9$X average) in success rate and up to $6$X ($2.7$X average) improvement in runtime over the current IBM Qiskit baseline. Our results also give insights to future system designers. Developing richer qubit topologies will reduce the need for SWAP operations and improve the reliability of important quantum primitives such as the Toffoli gate.


Our work is relevant for future QC systems for several reasons. Fundamental unreliability in qubits \cite{superconducting_stability} and short coherence times, even with Schoelkopf's coherence scaling law \cite{schoelkopflaw}, necessitate optimizations based on error rates and gate times. Although QEC is promising in the long run, even a single logical error-corrected qubit will be composed of many noisy qubits and our methods will be useful to perform noise-adaptive compilation of error correcting circuits. Our methods can also be extended to map programs to logical qubits based on their error properties. Our techniques can be adapted for other qubit technologies such as trapped ions \cite{trappedion3} and other routing approaches such as teleportation-based communication \cite{teleportation} by choosing the appropriate constraints in the optimization.

Overall, given the challenges of building reliable and scalable QC hardware, the key for the next five years or more will lie in ultra-efficient use of the resources available in NISQ systems. Our tool offers important leverage in stewarding runtime resource usage and optimizing reliability.  
